\documentclass[runningheads, 12pt]{llncs}

\usepackage{geometry}
\usepackage[colorlinks]{hyperref}
\hypersetup{
    colorlinks=true,
    linkcolor=red,   
    citecolor=blue,
    urlcolor=cyan,
    }
    
\usepackage{breakcites}
\usepackage{tikz-cd} 
\usepackage{circuitikz-0.8.3} 
\usepackage{makecell}
\usepackage{graphicx}
\usepackage{amssymb}
\usepackage{amsfonts} 
\usepackage{mathtools} 
\usepackage{dsfont} 
\usepackage{physics} 
\usepackage{todonotes}
\presetkeys{todonotes}{inline}{}
\usepackage{url} 
\newcommand\bydef{\stackrel{\mathclap{\normalfont\mbox{\normalfont\tiny def}}}{=}}
\def\eye{\mathds{1}} 

\def\ketbra#1#2{{\vert#1\rangle\!\langle#2\vert}}

\graphicspath{{figures/}}

\begin{document}
\title{The ZX-Calculus is Canonical in the Heisenberg Picture for Stabilizer Quantum Mechanics}

\author{J Biamonte\inst{1} \and A Nasrallah\inst{2}}

\authorrunning{Biamonte \and Nasrallah}

\institute{Yanqi Lake Beijing Institute of Mathematical Sciences and Applications\\Yanqi Island, Beijing, China 101408
\email{jacob.biamonte@deepquantum.ai}
\and 
Skolkovo Institute of Science and Technology\\ 
3 Nobel Street, Moscow 143026, Russian Federation
\email{aly.nasrallah@deepquantum.ai}
}
\maketitle              

\begin{abstract}
In 2008 Coecke and Duncan proposed the graphical ZX-calculus rewrite system which came to formalize reasoning with quantum circuits, measurements and quantum states.  The ZX-calculus is sound for qubit quantum mechanics.  Hence, equality of diagrams under ZX-equivalent transformations lifts to an equality of corresponding equations over matrices.  Conversely, in 2014 Backens proved completeness, establishing that any derivation done in stabilizer quantum mechanics with matrices can be derived graphically using the ZX-calculus. A graphical rewrite system that is both confluent and also terminates uniquely is called canonical.  Applying alternate sequences of rewrites to the same initial diagram, a rewrite system is confluent whenever all resulting diagrams can be manipulated to establish graphical equivalence. Here we show that a reduced ZX-rewrite system is already confluent in the Heisenberg picture for stabilizer quantum mechanics.  Moreover, any application of a subset of ZX-rewrites terminates uniquely and irrespective of the order of term rewrites in the Heisenberg picture for stabilizer quantum mechanics.  The ZX-system is hence Heisenberg-canonical for stabiliser quantum mechanics.   For a stabilizer circuit on $n$-qubits with $l$ single-qubit gates and $g$ two-qubit gates, the circuit output can be derived graphically in the Heisenberg picture using no more than $(\frac{1}{2}\cdot g+l)\cdot n$ graphical rewrites, thereby providing a graphical proof of the Gottesman-Knill theorem.  Finally, we consider the application of this tool as a graphical means to derive parent Hamiltonians which can be used as penalty functions in variational quantum computation.  Hence, we establish that each stabilizer state described by a Clifford circuit gives rise to a non-negative parent Hamiltonian with $n+1$ terms and a one-dimensional kernel spanned by the corresponding stabilizer state.  Such parent Hamiltonians can be derived with $\mathcal{O}(t\cdot n)$ graphical rewrites for a low energy state prepared by a $t$-gate Clifford circuit.  

\end{abstract}

\newpage 

Graphical reasoning has remained part of quantum information science since the early days of the field \cite{Feynman1986,godin1986quantum,haus1987quantum,Deutsch73}. More recently, the graphical language \cite{penrose1971applications} has become central to tensor network methods~\cite{biamonte2019lectures,biamonte2017tensor,Eis13,BC17,Oru14,CV09} whereas the formalisation of systematic graphical reasoning systems are now widely considered in research domains related to categorical quantum mechanics~\cite{AC04,heunen_categories_2019} as well as areas related to tensor networks~\cite{biamonte2019lectures,biamonte2017tensor,Eis13,BC17,Oru14,CV09}.  

One contemporary approach to graphical reasoning is the so called, Coecke-Duncan ZX-calculus \cite{CD,redgreen}, which formalizes graphical reasoning of typical quantum circuits, measurements and states appearing in quantum information processing \cite{CD,redgreen,WPB20,Backens_2014,Kissinger2015,TFY22,Bac14,JPV18}. The ZX-calculus has been used as a tool to study a range of tasks related to quantum computing e.g.~quantum circuit optimization \cite{Bea+20,Cow+20,KW20}, quantum circuit equivalence checking \cite{CSD20,PBW22} and in the design of quantum error correcting codes \cite{CHS19,CKA16}.  Diagrammatic manipulations using the ZX-calculus have also been automated by Kissinger and others \cite{Kissinger2015,TFY22}.

A graphical system is given by an axiomatic collection of graphical identities between diagrams. Each diagram is called a term whereas the fundamental identities are called, one-step equivalency transformations.   Typically in quantum circuits, rewrites are symmetric (aka reversible) and hence an equality symbol is used to denote the graphical equivalence relating terms. Non-symmetric rewrite systems are also possible. For example, one would replace term equalities with one-step arrows: a restriction we make in our main proof.   

Let $S$ be a set of terms (diagrams) and let $\rightarrow$ be the set of  one-step equivalency transforms on $S$. For $A\in S$ we denote by $\{A\rightarrow\}$ the set of diagrams in $S$ resulting from single step transformations on $A$.   We will now consider the equivalency class (aka closure) of $\rightarrow$: 

\begin{definition}[Arrow closures]
We denote by $\xrightarrow{\star}$ the reflexive, transitive closure of $\rightarrow$.
\end{definition}

Hence, we write $A\xrightarrow{\star} B$ to mean that there exists a sequence of single step transformations to arrive at $B$ starting from $A$.  
Now we will consider the closure of terms:   

\begin{definition}[$\star$-closure]
The reflexive, transitive $\star$-closure of term $A$ in a set of diagrams $S$ is given as: 
\begin{equation}
    \{A\xrightarrow{\star}\}\equiv\{B \in S | A\rightarrow B\}.
\end{equation}
\end{definition}
By reflexive we mean that there exists an identity on all terms $A\rightarrow A$.  Transitive means that $A\rightarrow B$ and $B \rightarrow C$ implies the existence of a map $A \rightarrow C$.  

\begin{definition}[Symmetric $\star$-closure]
The reflexive, transitive symmetric $\star$-closure of term $A$ in a set of diagrams $S$ is given as: 
\begin{equation}
   \{A\xleftrightarrow{\star}\}\equiv\{B \in S | A\leftrightarrow B\}.
\end{equation}
\end{definition}
Here symmetric means that all maps have an inverse.  Namely $A \rightarrow B$ implies a map $B \rightarrow A$. 

Given a ZX-diagram $A$, we denote the corresponding matrix representative as $[A]$, called the standard interpretation or flattening of $A$.  We say that a diagram $B$ is derivable from $A$ only when ZX-equivalent transformations can be used to manipulate $A$ into $B$.  One would then write $A\xrightarrow{\star} B$.  In other words, the $\star$-closure of $A$ contains a transformation(s) which derives $B$.  As the ZX-system is sound, this implies equality of the underlying matrices $[A]=[B]$.  More formally: 

\begin{definition}[Soundness]
Let $A$ be a ZX-diagram, then 
\begin{equation}
    \forall ~Q \in \{A\xleftrightarrow{\star}\} \implies ![R],    
\end{equation}
where $[R]$ is the flattening of any diagram in $\{A\xleftrightarrow{\star}\}$. 
\end{definition}

Hence, all diagrams in $\{A\xleftrightarrow{\star}\}$ correspond to a unique matrix $[R]$ given by the flattening of any diagram in $\{A\xleftrightarrow{\star}\}$ and called the symmetric $\star$-closure invariant of $A$.  This is soundness.  Note that two $\star$-closures need not have different $\star$-invariants.   For uniqueness in the case of stabiliser quantum mechanics, we need to recall the theory of Backens \cite{Backens_2014}.  Indeed, completeness is defined as the converse to soundness: if two-well formed equations can be proven to be equal using equations over matrices, then their diagrams related by ZX-equivalent transformations.  More formally: 

\begin{definition}[Completeness---Backens \cite{Backens_2014}]
Assuming stabilizer quantum mechanics, given a well formed equation deriving matrix $[A]$: 
\begin{equation}
    [A] \implies ! \{A\xleftrightarrow{\star}\}.
\end{equation}
\end{definition}

As mentioned in the abstract, the ZX-calculus is sound for qubit quantum mechanics \cite{CD,redgreen}, and complete for stabilizer quantum mechanics \cite{Backens_2014}. Note that the union of soundness and completeness induces a grading on the ZX-system.  The $\star$-closures are mutually disjoint and hence, the disjoint union of all $\star$-closures is 1-1 with the set of all ZX-diagrams.  This is what Backens proved \cite{Backens_2014}.
(See also the pseudo normal from of stabilizer ZX-diagrams given by Duncan and Pedrix in \cite{DP09}). 



Backens theory was adapted by Jeandel, Perdrix and Vilmart who showed completeness of a ZX-type calculus for graphically representing real valued matrices \cite{JPV17}.   Duncan and Pedrix also proved completeness in their formulation of real-valued stabiliser quantum mechanics \cite{DP13}. Approaches related to the extension of these results include universal extension of stabiliser ZX-calculus~\cite{CD,redgreen,NW17,Bac14,JPV18}.  Whereas Backens showed completeness of the ZX-calculus for single qubit Clifford gates plus $\pi/4$ T-gates \cite{Bac14} and a two-qubit Clifford+T extension was considered by Coecke and Wang \cite{CW18},  incompleteness of the ZX-calculus for the Clifford+T quantum mechanics was proven in \cite{Jea+17}.  Earlier incompleteness results for the ZX-calculus can be found in~\cite{WZ14}. 

In this work, we will restrict ourselves to the Heisenberg picture of stabilizer quantum mechanics.  Whereas in the standard setting of ZX-diagrams, the $\star$-closures partition the set of terms into mutually disjoint equivalency classes, in the Heisenberg picture for stabilizer quantum mechanics, there is indeed a unique normal form.  There are several definitions appear in the literature related to rewrite termination \cite{Der87}. We will firstly consider a weak form of termination.  

\begin{definition}[Weak termination]
$\forall A \in S$ there exists a $D \in S$ such that $D\in \{A\xrightarrow{\star}\}$ and $D= \{D\xrightarrow{\star}\}$. 
\end{definition}

When we write $D= \{D\xrightarrow{\star}\}$ we mean that $D$ is exactly equal to its one element star closure.  We then call $D$ terminal.  Of course, we are dealing with a restriction of the ZX-rules.  To show that the system weakly terminates, it is enough to show that a rewrite sequence exists which terminates uniquely.

\begin{definition}[Efficient termination]
Let $A$ be a $t$-term Heisenberg stabilizer ZX-diagram on $n$-qubits and let $D$ note $A$'s terminal element, termination is efficient whenever there exist a sequence of rewrites $A\xrightarrow{\star} D$ in $\mathcal{O}(\textrm{poly}(t\cdot n))$ steps.
\end{definition}

To prove confluence of the ZX-system in the Heisenberg picture for stabiliser quantum mechanics, we considered the property of weak termination of a subset of the ZX-rewrite system.  Informally this shows that there exists a unique terminal diagram that results from the application of a subset of the ZX-rewrite rules, independent of the order.  

\begin{definition}[Confluence]
For all $A, B, C \in S$ such that $A\xrightarrow{\star} B, C$, there exist a $D \in S$ such that $B, C\xrightarrow{\star} D=\{D\xrightarrow{\star}\}$.
\end{definition}

Hence, we show that a subset of equivalence transforms partitions 
all ZX-stabiliser diagrams disjointly into directed graphs.  As we establish that the stabilizer ZX-calculus is already confluent and (weakly) terminal in the Heisenberg picture, we show this by proving that any application irrespective of the order of a restricted form of rewrites results in a unique term. From these results, we establish that the rewrite system is already canonical:  
\begin{definition}[Canonical rewrite system \cite{Hul80}]\label{def:canonical}
A rewrite system is equivalently, 
\begin{enumerate}
    \item[(i)] canonical (aka convergent),
    \item[(ii)] weak-terminal and confluent.
\end{enumerate}
\end{definition}
Consequentially, confluence and terminality in the Heisenberg picture lead to a graphical variant of the Gottesman-Knill theorem \cite{stabs}. 

We continue by recalling the standard mathematical framework of qubit quantum mechancis in \S~\ref{sec:mathfram}. \S~\ref{sec:Clifford} presents a summary of Clifford gates and their properties: we recall the building blocks and the notation for the ZX-calculus in \S~\ref{sec:tnbb}. Note that whenever possible we tailor this notation to match quantum circuits.  In \S~\ref{sec:stabs} we recall the axiomatic rules for the ZX-rewrite system and we derive the rules we will use in the Heisenberg picture. We present our main results in \S~\ref{sec:confluence}: namely that the stabilizer ZX-calculus is confluent, terminal and hence canonical in the Heisenberg picture.  Finally, we show that these results allow one to recover the Gottesman-Knill theorem graphically. Then we apply these tools to graphically derive parent Hamiltonians for stabiliser states.  These can serve as penalty functions for variational quantum computation.  

\section{Standard mathematical framework of qubit quantum mechanics}\label{sec:mathfram}

The graphical ZX-system would replace and augment the standard matrix-based approach to describe quantum information processing.  As such, we will recall the standard definitions of qubit quantum mechanics.

We consider quantum computation with collections of two-level quantum systems acted on by unitary quantum gates.  

\begin{definition}[Qubit state space]
The $n$-qubit state space is defined as the complex vector space $V_n= [\mathbb{C}^2]^{\otimes n} \cong [\mathbb{C}]^{2^n}$ where a qubit state is a normalized vector in this space. 
\end{definition}
We will make use of the following conventions: 

\begin{enumerate}
    \item[(i)] We let $\ket{\psi} \in V_n$ represent a quantum state.
    \item[(ii)] We let $\bra{\psi} \in V_n^*$, where $(\ket{\psi})^\dagger = \bra{\psi}$ represent a costate or effect.
    
    \item[(iii)] We denote by $\mathcal{L}(V_n)$ the space of linear maps from $[\mathbb{C}^2]^{\otimes n}$ to itself.

    \item[(iv)] Quantum gates are given by unitary operators: 
    \begin{equation}
        U \in \mathcal{U}_{\mathbb{C}}(2^n) \equiv \left\{ U \in \mathcal{L}(V_n) | UU^\dagger = \eye \right\}. 
    \end{equation}
    

    \item[(v)] Hamiltonians are given as: 
    \begin{equation}
    \begin{aligned}
        A \in \textrm{herm}_{\mathbb{C}}(2^n) &\equiv \left\{ A \in \mathcal{L}(V_n) | A=A^\dagger \right\}\\
        &= \textrm{span}_{\mathbb{R}}\left\{\bigotimes_{i=1}^n \sigma_i^{\alpha_i} ~|~ \alpha_i = 0, 1, 2, 3 \right\}. 
         \end{aligned}
    \end{equation}
    \item[(vi)] The n-qubit basis is defined as $\mathcal{B}_n = \{\ket{0}, \ket{1}\}^{\otimes n}$, $\textrm{span}\{\mathcal{B}_n \}\cong V_n$, where $\ket{0} = (1, 0)^\top$ and $\ket{1} = (0, 1)^\top$. 
    \item[(vii)] Single qubit basis vectors are given as $\ket{0}, \ket{1} \in \mathcal{B}_1$, with $\textrm{span}\{\ket{0},\ket{1}\} \cong V_1$.
    \item[(viii)] The $2^n$ basis vectors are orthonormal and satisfy, $\braket{l}{m} = \delta_{l m}$.
\end{enumerate}

\begin{definition}[Pauli matrices]\label{def:pauli-basis}
Let us denote the Pauli matrices as follows:
\begin{equation}
\begin{aligned}
&\sigma^0\equiv\eye = \ketbra{0}{0} + \ketbra{1}{1}
,~~~\sigma^1\equiv X = \ketbra{0}{1} + \ketbra{1}{0},\\
&\sigma^2\equiv Y = \imath(\ketbra{1}{0}-\ketbra{0}{1})
,~~~\sigma^3\equiv Z =\ketbra{0}{0} - \ketbra{1}{1}.
\end{aligned}
\end{equation}
These form a basis for matrices in $Mat_{\mathbb{C}}(2)$.
\end{definition}

The Pauli $X$ matrix is sometimes called the NOT-gate it induces the mapping $\ket{0}\overset{X}{\longleftrightarrow}\ket{1}$.  The Pauli matrices satisfy 
\begin{equation}
    X^2=Y^2=Z^2=\eye,
\end{equation}
and 
\begin{equation}
    XYZ=\imath \eye. 
\end{equation}
Furthermore we have that: 

\begin{definition}[Pauli group]The Pauli matrices form a group over multiplication with elements: 
\begin{equation}
    \mathcal{P}_n= \left\{ e^{\imath \theta \frac{\pi}{2}}\bigotimes_{i=1}^n \sigma_i^{\alpha_i}~ |~ \theta, \alpha_i = 0, 1, 2, 3 \right\}. 
\end{equation}
\end{definition}

Informally, the normalizer of $\mathcal{P}_n$ is the set of unitary operations that leaves $\mathcal{P}_n$ invariant under element wise conjugation. These are known as Clifford gates.

\section{Clifford gates}\label{sec:Clifford}

The controlled-NOT (aka Feynman) gate is a central building block for quantum information processing.  The gate and its building blocks have been used extensively: it appears in classical~\cite{Lafont03towardsan} and quantum diagrammatic reasoning~\cite{CD,redgreen}, in categorical quantum mechanics \cite{CD,redgreen} and in tensor networks~\cite{2011AIPA....1d2172B,2011JPhA...44x5304B,2013JPhA...46U5301B,2012JPhA...45a5309D} and tensor contractions for \#SAT counting problems~\cite{2015JSP...160.1389B}.  

Let us recall the standard generating set of all Clifford group circuits. 

\begin{figure}
    \centering
    \includegraphics[width=\textwidth]{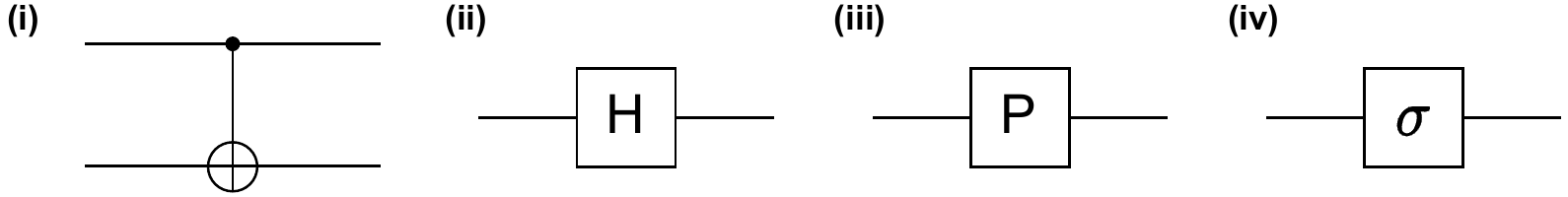}
    \caption{Clifford gates: (i) controlled-{X}, (ii) Hadamard, (iii) phase and (iv) arbitrary Pauli gates.}
    \label{fig:Cliffords}
\end{figure}

\begin{definition}[Clifford gates]
The following gates generate any Clifford circuit \cite{stabs,BA04}:
\begin{enumerate}
    \item[(i)] the controlled-{NOT} gate\footnote{The notation $C_m^l(X)$ is a controlled-X gate, where the $l^{th}$  and $m^{th}$ qubits are the control and target qubits respectively.}
    \begin{equation}
        C(X)=\ketbra{0}{0}\otimes \eye +\ketbra{1}{1}\otimes X; 
    \end{equation}
    
    \item[(ii)] the Hadamard gate 
    \begin{equation}
        H = \frac{1}{\sqrt{2}}(X+Z);  
    \end{equation}

    \item[(iii)] the phase gate 
    \begin{equation}
        P = \ketbra{0}{0} + \imath \ketbra{1}{1};      
    \end{equation}
    
    \item[(iv)] the Pauli gates and their products from Definition \ref{def:pauli-basis}. 
\end{enumerate}
\end{definition}

\begin{remark}
The gates (i-iv) are presented graphically in Figure \ref{fig:Cliffords}. These gates (i-iv) sequence to generate any Clifford circuit.  
\end{remark}

\begin{remark}
The standard properties of single qubit Clifford gates follow:  
\begin{enumerate}
 \item[(i)] $HXH=Z$ and $HZH = X$, 
 \item[(ii)] $PXP^\dagger = Y$ and $PZP^\dagger = Z = P^2$.
\end{enumerate}
\end{remark}

\begin{definition}[Clifford group]\label{def:Clifford}
The Clifford group is defined as the normalizer of the Pauli group as:
\begin{equation}
     \mathcal{C}_n=\left\{g \in \mathcal{U}(2^n)~|~ g p g^{\dagger} \in \mathcal{P}_n, \forall p \in \mathcal{P}_n \right\}.
\end{equation}
\end{definition}

We will now consider the controlled-NOT gate in further detail.  

\section{Tensor network building blocks and the ZX-caluclus}\label{sec:tnbb}

The building blocks of the controlled-NOT gate will now be separately considered. 
Firstly one will consider the black dot which copies binary inputs ($0$ and $1$) as: 
\begin{subequations}
\begin{align}
         0 &\rightarrow  0,0\label{eqn:copy}\\
         1 &\rightarrow  1,1.\label{eqn:copy2}
\end{align}
\end{subequations}
In the diagrammatic tensor network language, the COPY-gate is 
\begin{center}
\includegraphics[width=.15\textwidth]{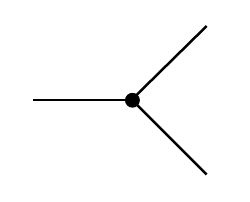}
\end{center} 
and graphically, equation \eqref{eqn:copy} and \eqref{eqn:copy2} become 
\begin{center}
\includegraphics[width=0.6\textwidth]{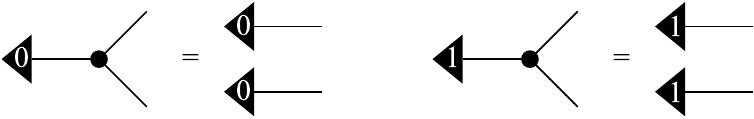}
\end{center} 
The next building block preforms the exclusive OR operation (XOR).  Given two binary inputs (say $a$ and $b$), the output ($a\oplus b = a + b - 2 ab$) is $1$ iff exactly a single input is $1$ (that is, addition modulo $  2$).  The gate is drawn as:
\begin{center}
\includegraphics[width=0.1\textwidth]{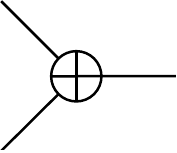}
\end{center} 
The XOR gate allows one to realize any linear Boolean function.  Let $f:\{0,1\}^n\rightarrow \{0,1\}$.  We consider indeterminates $x_1 x_2\dots x_n$.  Then $f(x_1,x_2,\dots,x_n)$ is linear over $\oplus$ if it can be written as:
\begin{equation}\label{eqn:linear}
  f = c_1 x_1 \oplus c_2 x_2 \oplus \cdots \oplus c_{n-1} x_{n-1} \oplus c_n x_n,
\end{equation}
where 
\begin{equation}
    {\bf c} \bydef (c_1, c_2, \dots ,c_{n-1}, c_n)
\end{equation}
is any $n$-long Boolean string.  Hence, there are $2^n$ linear Boolean functions and note that negation is not allowed.  When negation is allowed a constant $c_0\oplus f$ is added (mod $2$) to \eqref{eqn:linear} and for $c_0=1$, the function is called affine. In other words, negation is equivalent to allowing constant $1$ as:
\begin{equation}\label{eqn:constants}
\includegraphics[width=0.28\textwidth]{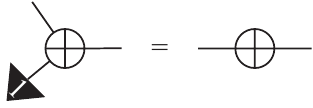}
\end{equation}
which sends Boolean variable $x$ to $1-x$.  Using the polarity representation of $f$, 
\begin{equation}
  \hat f ({\bf x}) = (-1)^{f({\bf x})}= 1-2f(\bf x).
\end{equation}
We note that linear Boolean functions index the columns of the $n$-fold tensor product of $2\times 2$ Hadamard matrices (that is, $H^{\otimes n}$ where the $i$--$j$th entry of each $2\times 2$ is $\sqrt{2}H_{ij} \bydef (-1)^{i\cdot j}$). In particular: 
\begin{equation}\label{eqn:xorH}
\includegraphics[width=0.3\textwidth]{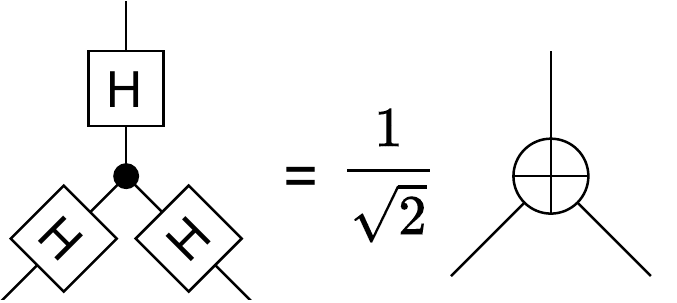}\tag{rule C1}
\end{equation} 
By \ref{eqn:xorH} one can think of XOR as being a copy operation in another basis.  We send binary $0$ to $\ket{0}$ and $1$ to $\ket{1}$. Then XOR acts as a copy operation: 
\begin{subequations}
\begin{align}
         \frac{1}{\sqrt{2}}\ket{+} &\rightarrow \ket{+,+},\label{eqn:xcopy}\\
         \frac{1}{\sqrt{2}}\ket{-} &\rightarrow \ket{-,-},\label{eqn:xcopy2}
\end{align}
\end{subequations}
using $H^2 = \eye$, $\ket{+} \bydef H\ket{0}$ and $\ket{-} \bydef H\ket{1}$.

A simplistic methodology to connect quantum circuits with indexed tensor networks starts with defining of two tensors in terms of components. 
\begin{center}
\includegraphics[width=0.45\textwidth]{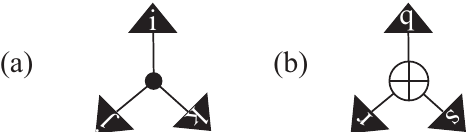}
\end{center} 
Using the Sengupta-Biamonte form \cite{SB22}, for (a) we have,
\begin{equation}
\begin{aligned}
  \delta^i_{~jk}&=\frac{1}{2} (i+j+k)^2 -\frac{3}{2} (i+j+k) + 1\\
 & = \frac{1}{2} (i + j + k -2) (i + j + k -1),
 \end{aligned}
\end{equation} 
where the indices $i$, $j$, $k$ $\in \{0,1\}$. In other words, the following contractions evaluate to unity. 
\begin{center}
\includegraphics[width=0.42\textwidth]{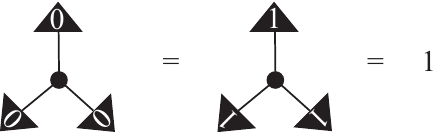}
\end{center} 
COPY tensor can be written as:
\begin{equation}
     \textrm{COPY} =\sum_{i,j,k} \delta^i_{~jk}  \ketbra{jk}{i}=\ketbra{00}{0} + \ketbra{11}{1}.
\end{equation}
Likewise using the Sengupta-Biamonte form \cite{SB22}, for (b) we have,
\begin{equation} 
\begin{aligned}
  [\oplus]^q_{~r s} &=
   -\frac{2}{3}(q+r+s)^3+3(q+r+s)^2 -\frac{10}{3}(q+r+s)+1 \\
  &= -\frac{2}{3}(q+r+s-3)(q+r+s-1)(q+r+s-\frac{1}{2}).
  \end{aligned}
\end{equation} 
Where the following contractions evaluate to unity (the XOR tensor is fully symmetric, hence the three rightmost contractions are identical by wire permutation). 
\begin{center}
\includegraphics[width=0.8\textwidth]{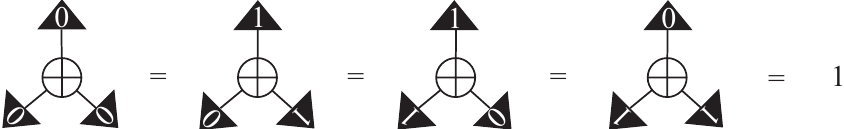}
\end{center} 
The XOR tensor can be expressed as:
\begin{equation}
       \textrm{XOR}=\sum_{q,r,s} [\oplus]^q_{~r s}  \ketbra{q}{rs}=
       \ket{0}(\bra{00}+\bra{11}) + \ket{1}(\bra{01}+\bra{10}). 
\end{equation}
Then the Feynman gate ($C(X)$) is given as the following tensor contraction,
\begin{equation}\label{eqn:cnot}
  \sum_m\delta^{ij}_{~~m}[\oplus]_{~qr}^{m} = {\text {C(X)}}^{ij}_{qr},
\end{equation}
where we raised an index on $\delta$.
All quantum circuits can be broken into their building blocks and thought of as indexed tensor contractions in this way. Typically a tensor network is written in abstract index notation.  We are concerned with the case where such equations can be replaced with wire diagrams (ZX-wire diagrams) entirely.  

\subsection{ZX-calculus}

As mentioned in the introduction, the ZX-calculus has been used in a variety of areas e.g.~measurement based quantum computation and quantum circuit optimization \cite{Bea+20,Cow+20,KW20}, quantum circuit equality validation \cite{PBW22}, and quantum error correcting codes \cite{CHS19,CKA16}. ZX-diagrams are equipped with a set of graphical rules that allows the graphical reasoning of linear maps. 

The building blocks of the ZX-rewrite system \cite{CD,redgreen} are given as follows:
\begin{center}
  \begin{tabular}{|p{2.cm}|p{3.2cm}|p{7.3cm}|p{2.7cm}|}
  \hline
      \textbf{Name} & {\bf Diagrammatic} & {\bf Braket/Operator Notation}& \textbf{Abstract Index Notation} \\
      \hline
       Identity & \makecell{\includegraphics[width=0.11\textwidth]{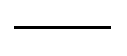}} & $  \eye = \ketbra{0}{0}+\ketbra{1}{1}$ & $ \delta^r_{s}$\\\hline
       Cup & \makecell{\includegraphics[width=0.11\textwidth]{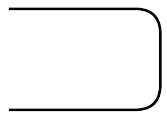}} & $  \sqrt{2}\ket{\phi^+}=(\ket{00}+\ket{11})$ & $ \delta^{rs}$\\ \hline
       Cap & \makecell{\includegraphics[width=0.11\textwidth]{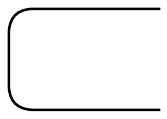}} & $  \sqrt{2}\bra{\phi^+}=(\bra{00}+\bra{11})$ & $ \delta_{rs}$\\ \hline
       Swap & \makecell{\includegraphics[width=0.11\textwidth]{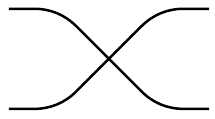}} & $\frac{1}{2}(\eye \otimes \eye + X \otimes X + Y \otimes Y +Z\otimes Z)$ & $ \delta^r_{p} \delta^s_{q}$\\ \hline
       Hadamard & \makecell{\includegraphics[width=0.11\textwidth]{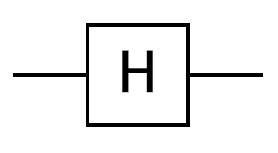}} & $  H= \ketbra{+}{0}+\ketbra{-}{1}$ & $  H^r_{s}$\\ \hline
       X vertices & \makecell{\includegraphics[width=0.21\textwidth]{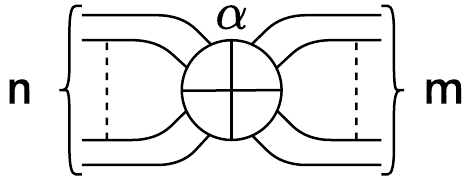}} & $  X^n_m(\alpha)=\ket{+}^{\otimes m} \bra{+}^{\otimes n}+ e^{\imath \alpha}\ket{-}^{\otimes m} \bra{-}^{\otimes n}$ & $  X^{r_1 r_2\cdots r_n}_{s_1 s_2\cdots s_m}(\alpha)$\\ \hline
       Z vertices & \makecell{\includegraphics[width=0.21\textwidth]{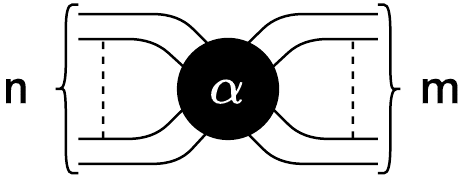}} & $  Z^n_m(\alpha)=\ket{0}^{\otimes m} \bra{0}^{\otimes n}+ e^{\imath \alpha}\ket{1}^{\otimes m} \bra{1}^{\otimes n}$& $  Z^{r_1 r_2\cdots r_n}_{s_1 s_2\cdots s_m}(\alpha)$\\ \hline
  \end{tabular} 
    \label{tab:ZX-gene}
\end{center}
where $\alpha= [0, 2 \pi)$, $ \ket{+}=\frac{1}{\sqrt{2}}(\ket{0}+\ket{1})$, and $ \ket{-}=\frac{1}{\sqrt{2}}(\ket{0}-\ket{1})$.

\begin{remark}\label{remark:notation}
One can derive the Pauli and Clifford gates from the table above:
\begin{enumerate}
    \item[(i)] The Pauli $Z$, which is recovered by setting ($n=1, m=1, \alpha = \pi$) in   $  Z^n_m(\alpha)$,
    \begin{equation}\label{eqn:Z}
        Z \equiv Z^1_1(\pi)=\ketbra{0}{0}-\ketbra{1}{1}. 
    \end{equation}

    \item[(ii)] The Pauli $  X$,  which is recovered by setting ($  n=1, m=1, \alpha = \pi$) in $  X^n_m(\alpha)$,
    \begin{equation}\label{eqn:X}
        X \equiv X^1_1(\pi)=\ketbra{+}{+}-\ketbra{-}{-}=\ketbra{0}{1}+\ketbra{1}{0}. 
    \end{equation}
    
    \item[(iii)] The Pauli $Y$, 
    \begin{equation}
        \imath Y= X^1_1(\pi)\cdot Z^1_1(\pi). 
    \end{equation}

    \item[(iv)] the Phase gate $P$,  which is recovered by setting ($n=1, m=1, \alpha = \frac{\pi}{2}$) in $  Z^n_m(\alpha)$,
    \begin{equation}
        P \equiv Z^1_1\left(\frac{\pi}{2}\right)=\ketbra{0}{0}+\imath\ketbra{1}{1}. 
    \end{equation}

    \item[(v)] The COPY tensor, which is recovered by setting ($  n=1, m=2, \alpha = 0$) in $Z^n_m(\alpha)$,
    \begin{equation}
        \textrm{COPY}  \equiv Z^1_2(0)=\ketbra{00}{0}+\ketbra{11}{1}. 
    \end{equation}

    \item[(vi)] The XOR tensor,  which is recovered by setting ($  n=2, m=1, \alpha = 0$) in $  X^n_m(\alpha)$,
    
    \begin{equation}
       \textrm{XOR}  \equiv X^2_1(0)=\ketbra{+}{++}+\ketbra{-}{--}. 
    \end{equation}

    \item[(vii)] The controlled-X gate, $C(X)$, which is the concatenation of COPY and XOR, 
    \begin{equation}
        C(X) = \ketbra{0}{0}\otimes \eye + \ketbra{1}{1}\otimes X. 
    \end{equation}

\end{enumerate}
\end{remark}

For single qubit gates we adopt the following graphical notations:

\begin{definition}[Z gate]
The dot on wire 
\begin{equation*}
  \includegraphics[width=.15\textwidth]{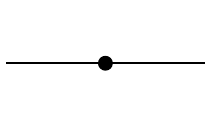}
\end{equation*}
denotes the $Z$ gate which correspond to \eqref{eqn:Z}. 
\end{definition}

\begin{definition}[X gate]
The plus on wire 
\begin{equation*}
  \includegraphics[width=.15\textwidth]{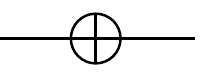}
\end{equation*}
denotes the $X$ gate which correspond to \eqref{eqn:X}.
\end{definition}
\begin{definition}[Y gate]
The Pauli $Y$ gate is represented graphically as a box with $Y$ in it. We have that $ZX= \imath Y$ which is represented graphically as:

\begin{equation*}
  \includegraphics[width=.4\textwidth]{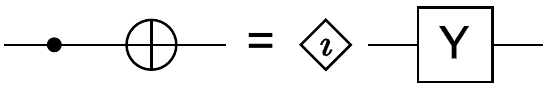}\tag{rule Y1}\label{rule:Y1}
\end{equation*}
and we also have $XZ= - \imath Y$ that follows graphically as:
\begin{equation*}
  \includegraphics[width=.4\textwidth]{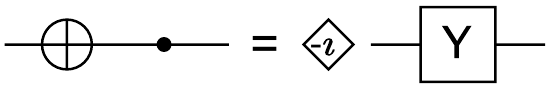}\tag{rule Y2}\label{rule:Y2}
\end{equation*}
where the $\imath, -\imath$ in the diamonds represents the complex scalars $\imath, -\imath$.
\end{definition}

The ZX-rewrite system considers the interact of the building blocks above.  This system subsumes the rewrites used in the present work.  However, we consider only stabiliser states and as such, only formulate the rules required to complete our proofs.  

\section{Stabilizer states}\label{sec:stabs}

We will now consider the basic definitions of stabiliser states.  Whereas the rewrite rules used here readily follow from (or are subsumed by) the ZX-rewrites, several of our rules also follow from the definition of stabilizers.

\begin{definition}[Stabilizer state]
An $n$-qubit state $\ket{\psi}$ is a stabilizer state of a subgroup $\mathcal{S} \subseteq \mathcal{P}_n$ if for all $P \in \mathcal{S}$, $P \ket{\psi}= (+1) \ket{\psi}$.
\end{definition}

\begin{example}[Single qubit stabilizers]
\
\begin{center}
  \begin{tabular}{| c | c | }
    \hline   
    $X$ & $\sqrt{2} \ket{+} = \ket{0} + \ket{1}$  \\ \hline
    $Y$ & $\sqrt{2} \ket{y^+} =\ket{0} + \imath \ket{1}$  \\ \hline
     $Z$ & $\ket{0}$  \\ \hline
     $-X$ & $\sqrt{2} \ket{-} = \ket{0} - \ket{1}$  \\ \hline
     $-Y$ & $\sqrt{2}\ket{y^{-}} = \ket{0} - \imath \ket{1}$  \\ \hline
     $-Z$ &  $\ket{1}$  \\ \hline
  \end{tabular}
\end{center}
\end{example}

\begin{example} 
The Bell state is a stabilizer state.
\begin{equation}
\sqrt{2} \ket{\phi^{+}} = \sum_{a,b} (a \oplus \lnot b  ) \ket{a,b} = \ket{00} + \ket{11},
\end{equation}
where $\lnot b = 1-b$. The abelian group that corresponds to this state is 
\begin{equation}
\mathcal{S} = \left\{\eye \otimes \eye, X\otimes X,-Y\otimes Y,Z \otimes Z\right\}.
\end{equation}
\end{example}

\subsection{Derivation of the rewrite rules}\label{sec:axioms}

The axiomatic rules of the ZX-rewrite system consists of a set of equations that describe how a diagram might be transformed into another. We will state some helpful graphical rules derived from the COPY and XOR tensors which will from some axiomatic rules for the rewrite system.

We will derive basic circuit identities and state them graphically, as is common in quantum circuits \cite{NC}.  Here we will adopt a tensor network approach and state identities that are also tensor symmetries \cite{biamonte2019lectures}. The rules presented in this section are used to derive templates in the Heisenberg picture, see \S~\ref{sec:templates}.

\begin{remark}
The COPY-tensor has stabilizer generators $X_iX_jX_k$, $Z_i Z_j$ for $  i\neq j\neq k \in \{1,2,3\}$ a qubit index. For example,  
\begin{equation}
     Z_iZ_j(\ket{000}+\ket{111})=\ket{000}+\ket{111}. 
\end{equation}
The following will present these identities graphically (Proposition \ref{prop:stabscopy}). 
\end{remark}

\begin{proposition}[Stabilizers of COPY]\label{prop:stabscopy}
The following equations are the Pauli stabilizers of COPY. 

\begin{equation}
     \includegraphics[]{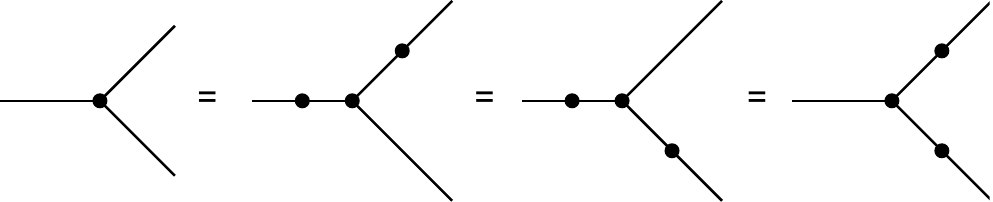}\tag{rule ST1}\label{rule:st1}
\end{equation}

\begin{equation}
     \includegraphics[]{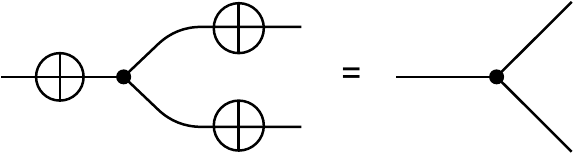}\tag{rule ST2}\label{rule:st2}
\end{equation}

\begin{equation}
     \includegraphics[]{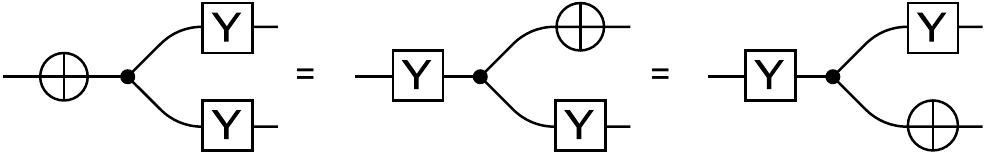}\tag{rule ST3}\label{rule:st3}
\end{equation}
\end{proposition}
Note that \ref{rule:st1} follows from spiders \cite{CD,redgreen}, \ref{rule:st2} follows from \ref{rule:K1a} below. \ref{rule:st3} are template identities that can be derived by substituting \ref{rule:st1} in \ref{rule:st2} and by using \eqref{rule:Y1} and \eqref{rule:Y2}.

Note that various other identities can be derived for stabilizer tensors, such as the following. Let $\ket{\psi}$ be the 3-qubit GHZ-state we have 
\begin{enumerate}
    \item[(i)] $Z_i\ket{\psi} = Z_j\ket{\psi}$, which follows graphically as:
\begin{equation*}
    \includegraphics[width=0.5\textwidth]{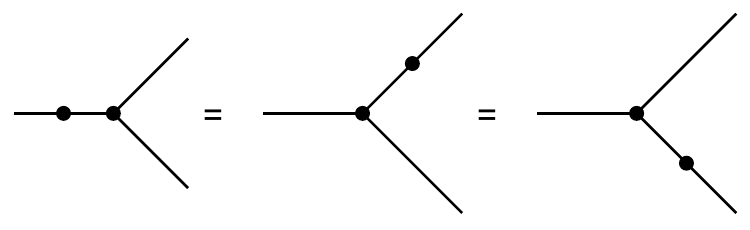}\tag{rule P1}
\end{equation*}
    \item[(ii)] $X_i\ket{\psi} = X_j X_k\ket{\psi}$, which follows graphically as:
\begin{equation*}
    \includegraphics[width=0.3\textwidth]{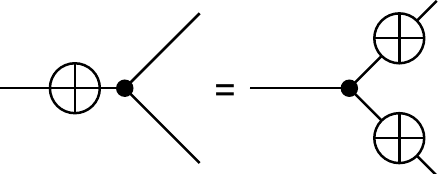}\tag{rule K1a}\label{rule:K1a}
\end{equation*}
\end{enumerate}

We use the XOR tensor to derive 2 more rules. Let $\ket{\phi}= \ket{000}+ \ket{011} + \ket{101}+ \ket{110}$ be the state that represents the XOR tensor with all wires bent to one direction, we have
\begin{enumerate}
    \item[(i)] $X_i \ket{\phi} = X_j \ket{\phi}$, graphically it's represented as:
    \begin{equation*}
        \includegraphics[width=0.5\textwidth]{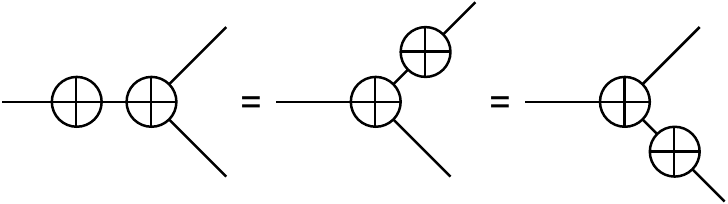}\tag{rule P2}
    \end{equation*}
    \item[(ii)] $Z_i\ket{\phi} = Z_j Z_k\ket{\phi}$, graphically it's represented as:
        \begin{equation*}
        \includegraphics[width=0.4\textwidth]{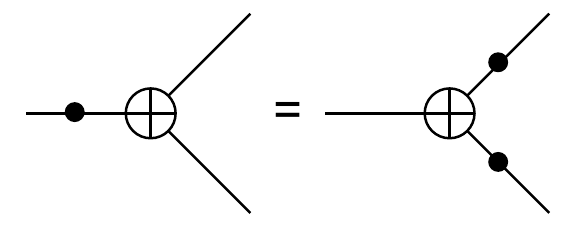}\tag{rule K1b}\label{rule:K1b}
    \end{equation*}
\end{enumerate}

\begin{proposition}[$Z^2=X^2=H^2=\eye$] These identities are given graphically as follows:
\begin{equation*}
    \includegraphics[]{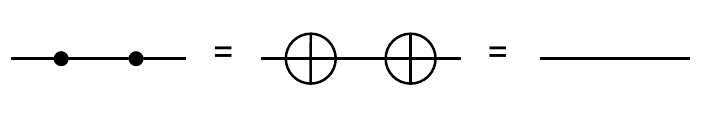}\tag{rule I1}
\end{equation*}

\begin{equation*}
    \includegraphics[]{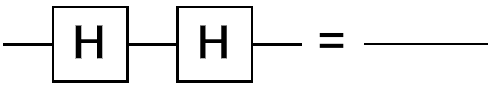}\tag{rule C3}\label{rule:C3}
\end{equation*}

\end{proposition}

\begin{definition}[Hopf law]\label{rule:hopf}
The Hopf law is defined graphically as follows~\cite{CD,redgreen}:

\begin{equation}
    \includegraphics[]{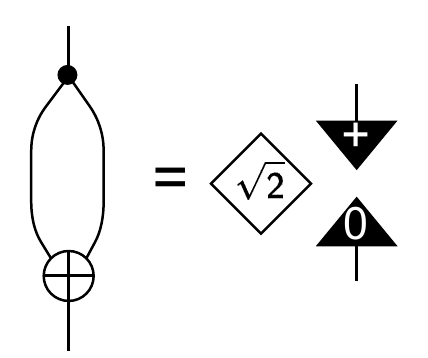}\label{eqn:hopf}
\end{equation}
\end{definition}

\begin{proposition}[$C(X)^2=\eye$]\label{prop:c21}
Using the Hopf law (rule \ref{eqn:hopf}) we derive graphically the identity ${C(X)} ^2=\eye$ as follow:
\begin{equation}
    \includegraphics[width = .9\textwidth]{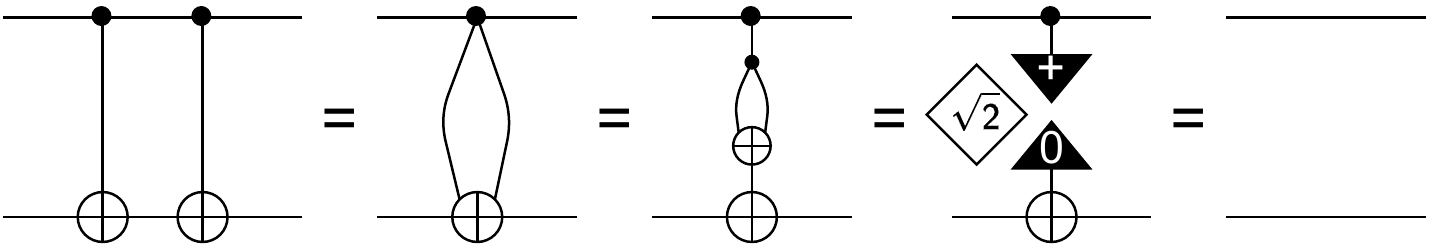}\tag{rule I2}
\end{equation}
\end{proposition}


\subsection{Heisenberg picture}

We consider the following unitary operator as:
\begin{equation}
  U_t = e^{-\imath t \mathcal{H}}, \hspace{30pt} U_t U_{t}^{\dagger} = \eye,
\end{equation}
with $\mathcal{H} \in herm_\mathbb{C}(2^n)$. The time evolution of a quantum state $\ket{\varphi_0}$ is 
\begin{equation}
  \ket{\varphi_t} = e^{-\imath t \mathcal{H}} \ket{\varphi_0}, \hspace{30pt} \forall t \braket{\varphi_t}{\varphi_t} =1. 
\end{equation}
This is called \text{Schr\"odinger's picture}. It results in gate sequences applied to input states such as $\prod_{l=1}^p U_l \ket{0}^{\otimes n}$, for $U_l \in \mathcal{U}_\mathbb{C}(2^n)$.

A second time evolution formalism is called \text{Heisenberg's picture}. Suppose we have a quantum system in state $\ket{\psi}$, we apply unitary $  U$ ($U U^{\dagger} = \eye$)
\begin{equation}
  UN\ket{\psi} = UNU^{\dagger}U \ket{\psi}.
\end{equation}
Then the evolution of operator $N$ is given by 
\begin{equation}
\label{eq:evoN}
  N \to U N U^{\dagger}.
\end{equation}

\begin{enumerate}
\item We want to follow the evolution of a number of $N$'s to reconstruct the evolution of $\ket{\psi}$.
\item Evolution in \eqref{eq:evoN} is linear so we will follow a complete basis of $n \times n$ matrices.
\end{enumerate} 

We call $  \mathcal{P}_n$ the Pauli group. As mentioned before it contains $  4 \cdot 4^n$ elements. These elements are tensor products of $  X, Y, Z, \eye$ and with prefactors $  \pm 1, \pm \imath$. There's a multiplicative group homomorphism 
$$MN \to UMN U^{\dagger} = (UMU^{\dagger})(UNU^{\dagger}),$$
so we can follow just a generating set of the group. A good one for the Pauli group is $\{X_1, ..., X_n, Z_1, ..., Z_n \}$.

The set of operators that leave $\mathcal{P}_n$ fixed under conjugation form the normalizer called the Clifford group $\mathcal{C}_n$. This group $\mathcal{C}_n$ is much smaller than the unitary group on $n$-qubits, $\mathcal{U}_{\mathbb{C}}(2^n)$, yet contains many operations of interest.

\begin{remark}
The transformation of a Pauli string $P\in \mathcal{P}_n$ in the Heisenberg picture as $P\rightarrow UPU^\dagger$ is restricted to unitaries from the Clifford group $U\in \mathcal{C}_n$.
\end{remark}

\subsection{Heisenberg rules of the stabilizer ZX-calculus}\label{sec:templates}
We define a set of rules that we derive form the transformation of the Pauli group generators under Clifford gate conjugation. We shall see how the Pauli group generators,
\begin{equation}\label{eqn:systoev}
    \{Z_1, \ldots Z_n, X_1, \ldots X_n\},
\end{equation}
evolve under conjugation by $H$, $P$, $C(X)$. To simulate the evolution of \eqref{eqn:systoev} there are $2n$ tensor contractions.  Each can be done graphically.  We can establish the following:
\begin{equation}\label{eqn:xtozbasis}
  H\colon Z \rightarrow X,~~~~H\colon  X\rightarrow Z.
\end{equation}
\begin{equation}\label{eqn:phase-conjugation}
  P\colon  X \rightarrow Y,~~~~P\colon Z\rightarrow Z.
\end{equation}
\begin{equation}\label{XZconjCN}
\begin{split}
 C(X)\colon  \hspace{2pt} X\otimes \eye & \rightarrow X\otimes X, \hspace{27pt} (i)\\
 \eye \otimes X &  \rightarrow \eye \otimes X, \hspace{29pt} (ii)\\
 X\otimes X & \rightarrow X\otimes \eye, \hspace{27pt} (iii)\\
Z \otimes \eye & \rightarrow Z \otimes\eye, \hspace{30pt} (iv)\\
 \eye\otimes Z &  \rightarrow Z\otimes Z, \hspace{31pt} (v)\\
  Z\otimes Z & \rightarrow \eye\otimes Z, \hspace{31pt} (vi)\\
  Z \otimes X &  \rightarrow Z \otimes X, \hspace{26pt} (vii)\\
  X \otimes Z &  \rightarrow -Y \otimes Y. \hspace{19pt} (viii)
\end{split}
\end{equation}


\begin{proposition}[Computational and $\pm$-Basis Change] \label{prop:HZH}
The l.h.s~of \eqref{eqn:xtozbasis} is given graphically as follows: 
\begin{equation}
    \includegraphics[width=.45\textwidth]{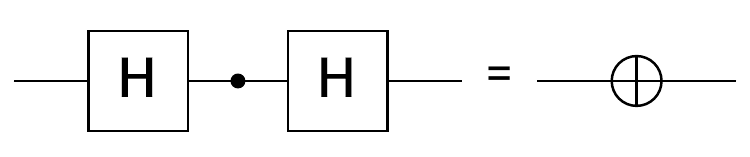}\tag{rule H1}
\end{equation}
\end{proposition}

\begin{remark}
In what follows, we denote by $\overset{(-)}{=}$ the one-step equivalency transform between two diagrams, where $(-)$ denotes the name of the rule applied.
\end{remark}

\begin{proposition}\label{prop:HXH}
From Proposition \ref{prop:HZH} and \ref{rule:C3} we recover the r.h.s~of \eqref{eqn:xtozbasis}.

\begin{equation}
    \includegraphics[width=.9\textwidth]{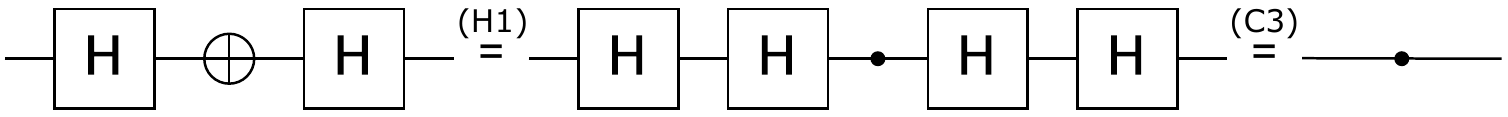}\tag{rule H2}
\end{equation}
\end{proposition}

\begin{proposition}\label{prop:PXP}
The l.h.s~of \eqref{eqn:phase-conjugation} is given graphically as:
\begin{equation}
        \includegraphics[width=.45\textwidth]{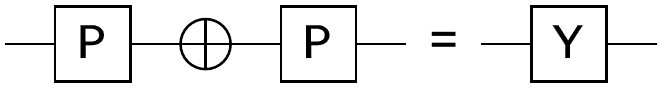}
\end{equation}
and the r.h.s~is given as:
\begin{equation}
        \includegraphics[width=.45\textwidth]{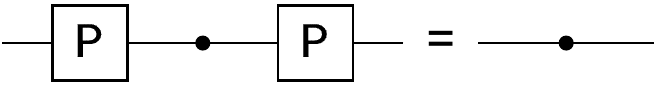}
\end{equation}
\end{proposition}

\begin{proposition}\label{prop:CNpCN}
The relationship \eqref{XZconjCN} is derived graphically as follow:
\paragraph{(i). $C(X)\colon X\otimes \eye  \rightarrow X\otimes X$}
\begin{equation}
    \includegraphics[width=\textwidth]{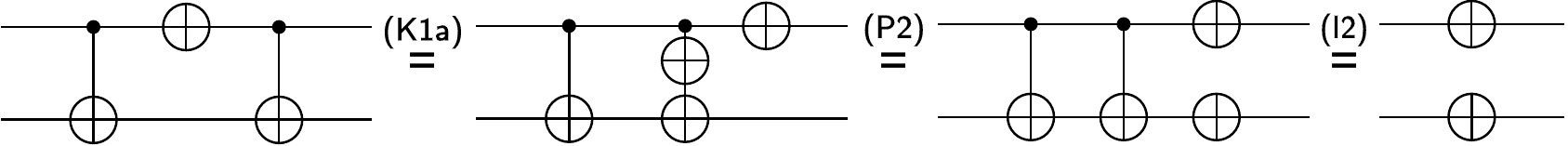}\tag{rule R1}
\end{equation}
\paragraph{(ii). $C(X)\colon \eye \otimes X  \rightarrow \eye \otimes X$}
\begin{equation}
    \includegraphics[width=.75\textwidth]{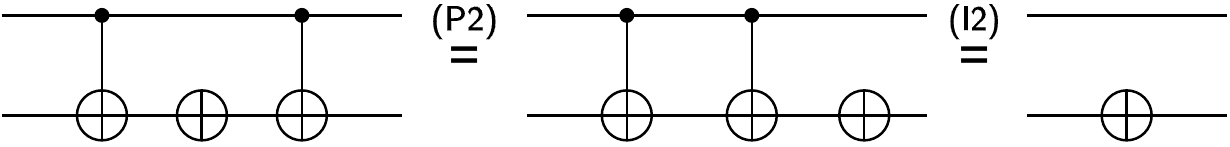}\tag{rule R2}
\end{equation}

\paragraph{(iii). $C(X)\colon X\otimes X  \rightarrow X\otimes \eye$}
\begin{equation}
    \includegraphics[width=.9\textwidth]{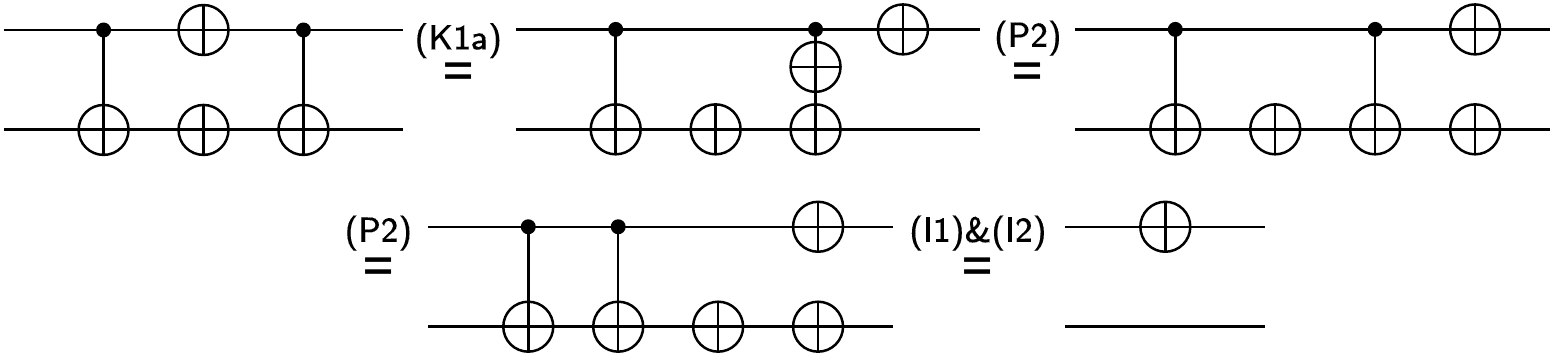}\tag{rule R3}
\end{equation}

\paragraph{(iv). $C(X)\colon Z \otimes \eye \rightarrow Z \otimes\eye$}
\begin{equation}
    \includegraphics[width=.75\textwidth]{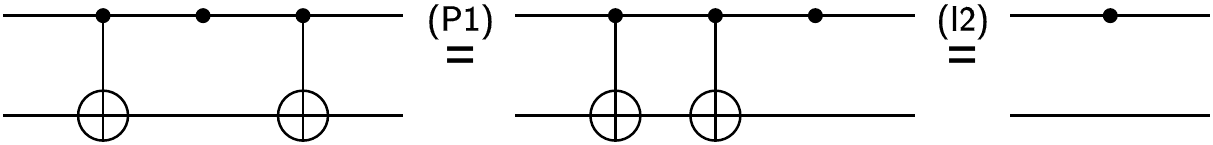}\tag{rule R4}
\end{equation}

\paragraph{(v). $C(X)\colon \eye\otimes Z \rightarrow Z\otimes Z$}
\begin{equation}
    \includegraphics[width=\textwidth]{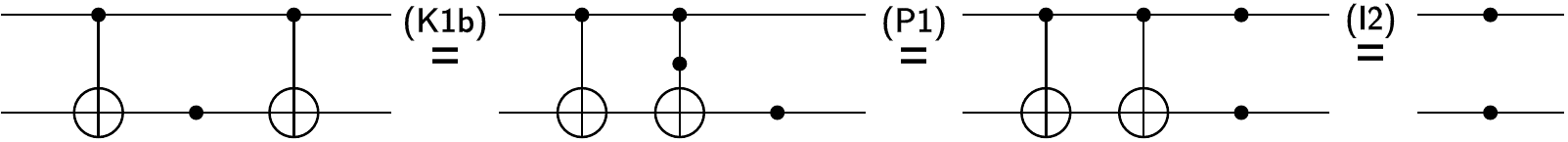}\tag{rule R5}
\end{equation}

\paragraph{(vi). $C(X)\colon Z\otimes Z  \rightarrow \eye\otimes Z$}
\begin{equation}
    \includegraphics[width=.9\textwidth]{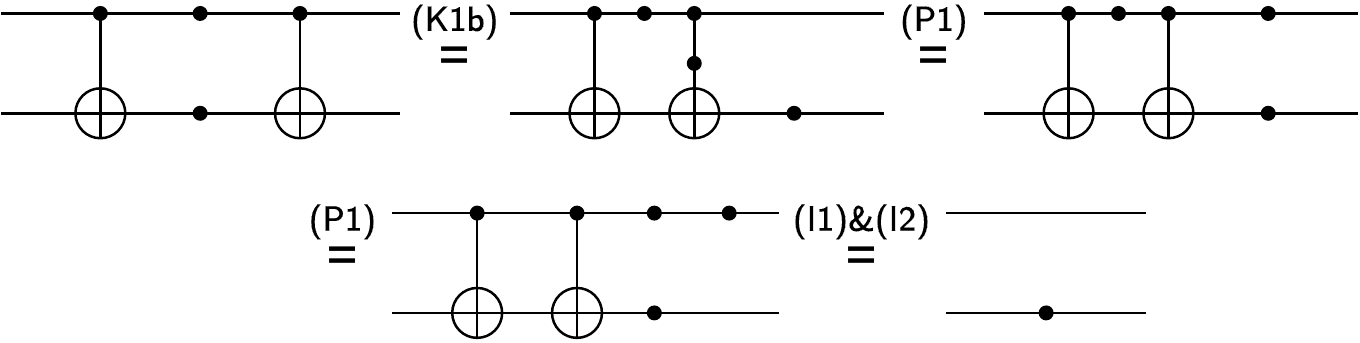}\tag{rule R6}
\end{equation}

\paragraph{(vii). $C(X)\colon Z \otimes X \rightarrow Z \otimes X$}
\begin{equation}
    \includegraphics[width=.9\textwidth]{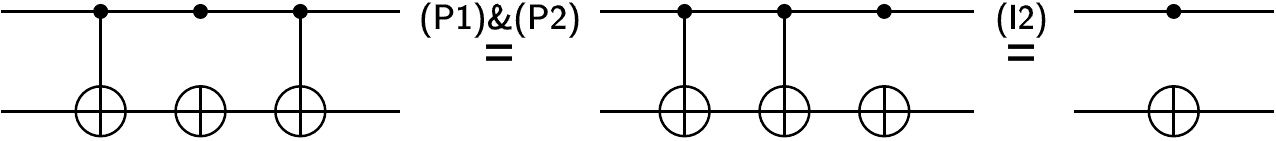}\tag{rule R7}
\end{equation}
 
\paragraph{(viii). $C(X)\colon X \otimes Z \rightarrow -Y \otimes Y$}
\begin{equation}
    \includegraphics[width=\textwidth]{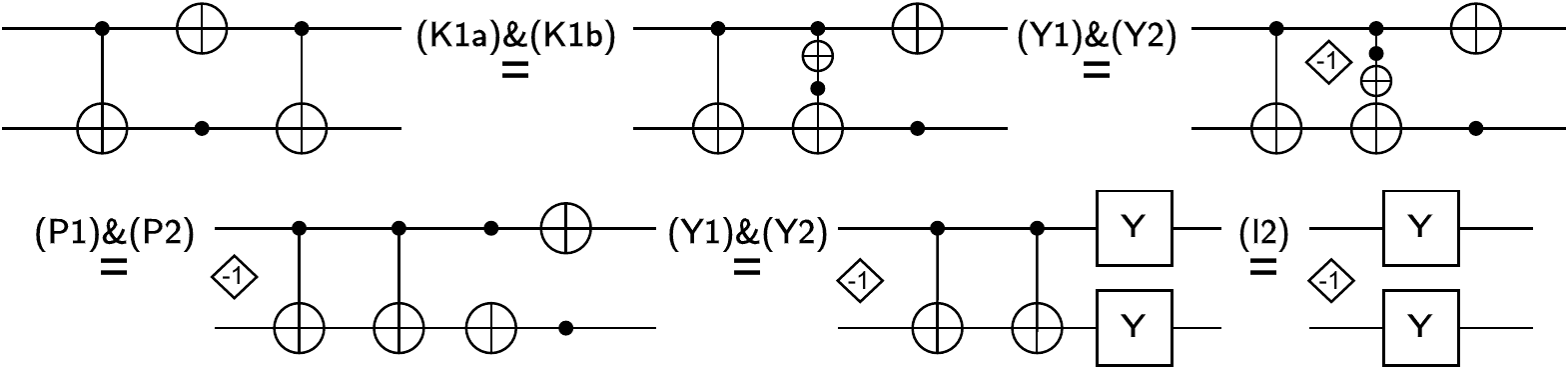}\label{eqn:CNonXZ}\tag{rule R8}
\end{equation}
where the $-1$ in the diamond represents the scalar $-1$.
\end{proposition}
\begin{remark}
The rules derived in the Heisenberg picture in this section and in Appendix \ref{appendix:Heisenberg-rules}, are called the Heisenberg rules.
\end{remark}
\begin{proposition}[Stabilizers Transform Covariantly]\label{prop:covariant}
Let $N$ be a stabilizer of the state $\ket{\psi}$, then $UNU^\dagger$ is a stabilizer of the state $U\ket{\psi}$.
\end{proposition}
\begin{proof}
$  \forall N \in S$ (set of stabilizers of the state $\ket{\psi}$) we have
\begin{equation*}
     N\ket{\psi}=\ket{\psi},
\end{equation*}
then
\begin{equation*}
     U\ket{\psi}=UN\ket{\psi}=UNU^\dagger~ U\ket{\psi},
\end{equation*}
which means $  UNU^\dagger$ is a stabilizer of $  U\ket{\psi}$. Hence, stabilizers transform covariantly $  N \rightarrow UNU^\dagger$. 
\end{proof}

\begin{example}
We will now consider an extended example related to Bell state stabilizers.  Let us now verify by tensor contraction that 
$$
  \{X\otimes X,-Y\otimes Y,Z\otimes Z\},
$$
are indeed stabilizers. The following circuit produces the Bell state $\ket{\phi^+}$. 

\begin{center} 
	\begin{circuitikz}
	\draw (1,1) -- (2,1);

	\draw (1,0) -- (2,0);
	\draw [fill] (2,1) circle 
[radius=0.1] ;
	\draw (2,1) -- (2,-0.2);
	\draw (2,0) circle [radius=0.2] ;

	\draw (2,1) -- (2.7,1);
	\draw (2.7,1.3) rectangle (3.3,0.7);
	\draw [scale=1.1](2.73, 0.93)  node [black] {$  H$};
	\draw (3.3,1) -- (4,1);

	\draw[fill] (4, 1.4) -- (4, 0.6) -- (4.5, 1) -- cycle;
	\draw (4.18, 1) node [white] {$  0$};

	\draw (2,0) -- (4,0);
	\draw[fill] (4, 0.4) -- (4, -0.4) -- (4.5, 0) -- cycle;
	\draw (4.18, 0) node [white] {$  0$};

\draw[dotted,rounded corners=2ex] (1.5,-0.5) rectangle (3.5,1.5);
\draw (2.5, -0.8) node {$  u$};

\draw (3, 2.5) node [black] {$\underset{\overbrace{\hphantom{qwertyuiopasdfg}}}{\ket{\phi^+}} $};

\draw (4.15, 1.7) node [black] {$\underset{\overbrace{\hphantom{qwer}}}{ \ket{00}} $};
\end{circuitikz}
\end{center} 
Such that,
\begin{equation}
     \ket{\phi^+} = \frac{\ket{00}+\ket{11}}{\sqrt[]{2}} = u\ket{00}. 
\end{equation}

We begin by application of Proposition \ref{prop:covariant}. First, consider stabilizers of the initial state, as follows. 

\begin{equation}
     Z\otimes Z\ket{00}=\ket{00},
\end{equation}

\begin{equation}
     Z\otimes \eye \ket{00}=\ket{00},
\end{equation}

\begin{equation}
     \eye\otimes Z\ket{00}=\ket{00}.
\end{equation}
Hence, the stabilizers of $ \ket{00}$ form the Abelian group \eqref{eqn:abgroup}. 
\begin{equation}\label{eqn:abgroup}
     \{\eye, Z\otimes \eye,\eye \otimes Z, Z\otimes Z\}.
\end{equation}
Acting on the initial state $  \ket{00}$ with $  u$ yields: 
\begin{equation}
\begin{split}
  u \ket{00} = \ket{\phi^+}, \\   N\ket{00} = \ket{00}. 
\end{split}
\end{equation}
From Proposition \ref{prop:covariant} we know that,
\begin{equation}
\begin{aligned}
  N\ket{00}&  =\ket{00},\\
  u\ket{00}&  =uNu^\dagger \ket{\phi^+}.
\end{aligned}
\end{equation}
Hence, Proposition \ref{prop:covariant} asserts that if \eqref{eqn:abgroup} is a stabilizer of the initial state, then \eqref{eqn:abgroupcov} is a stabilizer of the final state $u\ket{00}$.

\begin{equation}\label{eqn:abgroupcov}
     \left\{\eye, u (Z\otimes \eye) u^\dagger, u (\eye \otimes Z) u^\dagger, u (Z\otimes Z) u^\dagger\right\}.
\end{equation}
We then must determine e.g.~$(Z\otimes Z) u^\dagger = u^\dagger \sigma'$. This is done graphically as follows: 

\begin{equation*}
    \includegraphics[width=\textwidth]{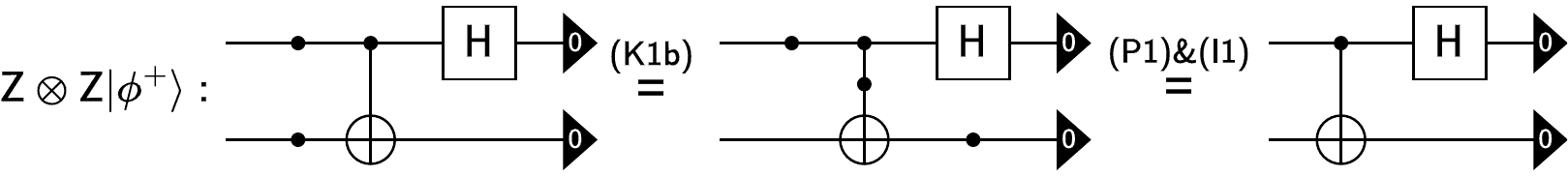}
\end{equation*}
Which simply shows that $Z\otimes Z$ is a stabilizer of $\ket{\phi^+}$, as expected. We consider now on recovering the stabilizer group \eqref{eqn:abgroupcov}. Consider then evolution of $Z\otimes \eye$.  We arrive at the following rewrites:
\begin{center}
    \includegraphics[width=.8\textwidth]{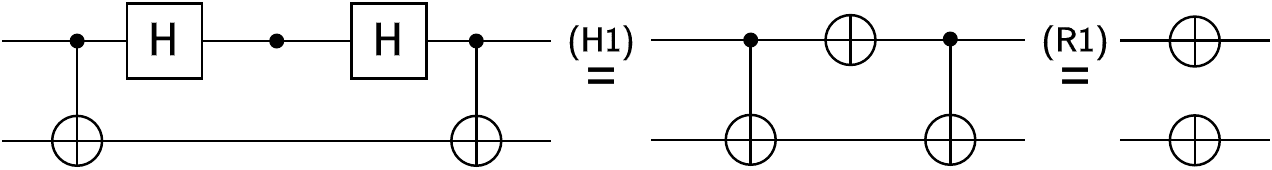}
\end{center}
Which arrives graphically at $  X\otimes X$ being a stabilizer of $\ket{\phi^+}$. We consider now the evolution of $\eye \otimes Z$, 
\begin{center}
    \includegraphics[width=.8\textwidth]{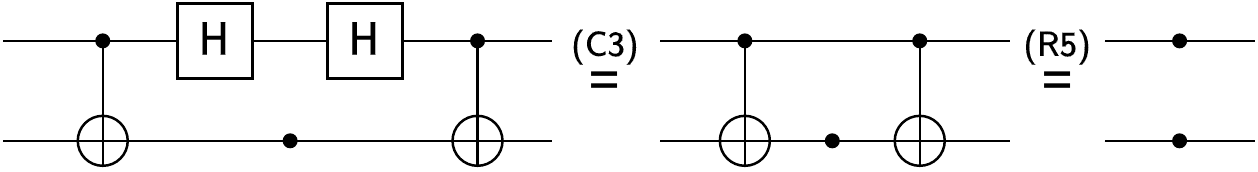}
\end{center}
Which arrives graphically at $  Z\otimes Z$ being a stabilizer of $\ket{\phi^+}$. Finally the evolution of $Z\otimes Z$, 
\begin{center}
    \includegraphics[width=1.\textwidth]{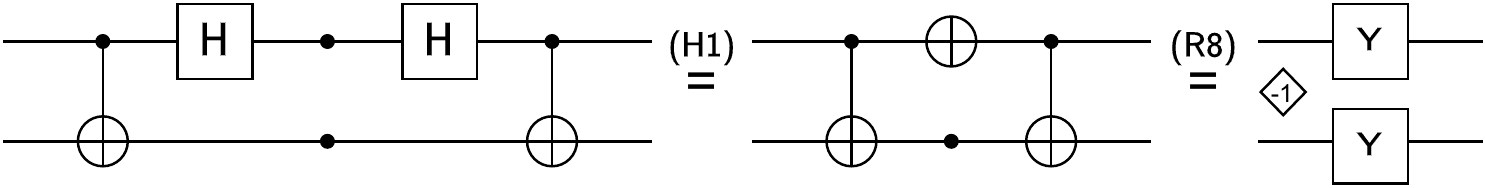}
\end{center}
Which arrives graphically at $-Y\otimes Y$ being a stabilizer of $\ket{\phi^+}$. Hence we recover the set of stabilizers of $\ket{\phi^+}$, $  S_{\ket{\phi^+}}=\{\eye, X\otimes X,-Y\otimes Y,Z\otimes Z\}$.
\end{example}

\section{Confluence and Termination in the Heisenberg picture}\label{sec:confluence}

In this section we show that the ZX-rewrite system is confluent, terminal and hence canonical in the Heisenberg picture for stabilizer quantum mechanics. As a corollary we present a graphical proof of the Gottesman-Knill theorem. We will show first that the Heisenberg stabilizer ZX-rewrites system is weakly terminal by showing the existence of a terminal element that is equal to its $\star$-closure for any term. Moreover, we show that reaching the terminal element using the Heisenberg rules would take at most $\mathcal{O}(\textrm{poly}(t))$ steps for a $t$-gate stabalizer evolution in the Heisenberg picture.  Furthermore, changing the order of rule applications does not modify the terminal element.  This unique path to a single terminating element implies that the Heisenberg stabilizer rewrite system is confluent. Hence the rewrite system is already canonical. 

Let $P, \Tilde{P}\in \mathcal{P}_n$ be two Pauli strings, and let $U\in \mathcal{C}_n$ be a Clifford circuit. We denote by $A$ the ZX-diagram representing $[UPU^\dagger]$, and by $D$ the ZX-diagram representing $[\Tilde{P}]$, such that $[UPU^\dagger]=[\Tilde{P}]$.

\begin{proposition}[Weak termination of the Heisenberg stabilizer ZX-rewrite system]\label{prop:weak-termination}
For every Pauli string mapped under Clifford conjugation ($A\in \{A\xrightarrow{\star}\}$) there exist a unique terminal element ($D=\{D\xrightarrow{\star}\}$) that is derivable using the rules of the rewrite system. 
\end{proposition}

\begin{proof}
The proof follows from noting that 
$D$ is derivable from $A$ under the Heisenberg rules, thus $D$ is in the $\star$-closure of $A$, $D\in \{A\xrightarrow{\star}\}$.
Moreover, a Pauli string is represented graphically as a countable collection of the diagrams representing the $\eye$, $X$, $Y$, and $Z$ gates, thus it is represented uniquely in the rewrite system. Hence, $D$ is equal to its $\star$-closure ($D=\{D\xrightarrow{\star}\}$), as there is no Heisenberg rules that reduces $D$ any further.   
\end{proof}

\begin{proposition}[Graphical rewrite upper bound]\label{prop:upper-bound}
Given a Pauli-string under Clifford conjugation of an $n$-qubit circuit with $t$-gates, the rewrite system terminates producing a Pauli-string in $\mathcal{O}\left(t\cdot n\right)$ steps.
\end{proposition}

To establish Proposition \ref{prop:upper-bound}, we consider two cases.  The first is for single-qubit Clifford gates acting by conjugation on Pauli operators. The second is the case of the 2-qubit Clifford gate ($C(X)$), acting by conjugation on 2-qubit Pauli strings. 

\begin{proof}
First case:
$U$ only contains a single 1-qubit Clifford gate. The only Pauli terms that will be affected are the terms where $U$ is positioned.
Using the rules defined in \S~\ref{sec:templates}, the number of graphical rewrites in this case is $1$, moreover if there is $l$ gates in the circuit the number of rewrites is $l$. Hence if the Clifford circuit is acting on $n$ qubits with $l$ 1-qubit gates, the number of graphical rewrites is upper bounded by $(l\cdot n)$.

Second case:
$U$ has only the 2-qubit Clifford gate.
Using the rules derived in \S~\ref{sec:templates} and Appendix \ref{appendix:Heisenberg-rules}, the number of graphical rewrites is 1. 
For $g$ gates in the circuit, the number of graphical rewrites will be $g$. Hence, for a Clifford circuit acting on $n$ qubits with $g$ 2-qubit gates, the number of graphical rewrites is upper bounded by $\frac{1}{2}\cdot g\cdot n$. This establishes the desired upper bound as, $(\frac{1}{2}\cdot g+l)\cdot n$.
\end{proof}

Proposition \ref{prop:weak-termination} established that for a given Pauli string under Clifford conjugation there exist a unique terminal element that is equal to its own $\star$-closure. 

\begin{remark}
Ambiguity in the graphical calculation shows  that there exist different sequences of Heisenberg rule applications that converges to the same terminal element, which implies that the system is confluent. It follows that, the Heisenberg stabilizer ZX-rewrite system is already canonical.
\end{remark}

Irrespective of the order of Heisenberg rule application, the system converges to the unique terminal element, hence: 

\begin{proposition}[Confluence of the Heisenberg stabilizer ZX-rewrite system]
The stabilizer ZX-rewrite system is confluent in the Heisenberg picture.
\end{proposition}

The Gottesman-Knill follows directly as in \S~\ref{sec:templates} we showed how the generators of the Pauli group are mapped graphically under Clifford conjugation, and that irrespective of the path of graphical calculation the number of rewrites is upper bounded by $(\frac{1}{2}\cdot g+l)\cdot n$.

\begin{corollary}[Graphical Proof of the Gottesman--Knill Theorem] \label{theorem:gk}
The Heisenberg evolution of an initial state $\ket{0^n}$ acted on by a $t$-gate Clifford circuit is determined graphically by not more than $(\frac{1}{2}\cdot g+l)\cdot n$ rewrites, thereby recovering the Gottesman-Knill theorem \cite{stabs}. 
\end{corollary}

\section{Applications to Penalty Functions}

In this section we aim to use the stabilizer ZX-calculus as a tool to derive Hamiltonian penalty functions. We will make use of the telescoping construction \cite{Bia21} to retrieve the Hamiltonian that has as its lowest energy state a stabiliser state. A telescoping construction is a Hamiltonian that encodes as its ground state the output state of a given quantum circuit. As an application, we retrieve the parent Hamiltonian of $n$-qubits Greenberger–Horne–Zeilinger state (GHZ state). The telescoping construction reads as:

\begin{definition}[Telescopic construction \cite{Bia21}]\label{def:telescopes}
For a given quantum circuit $U$, the non-negative telescoping Hamiltonian is written as $H_{\textrm{teles}}= UP_0 U^\dagger \geq 0$ where $P_0$ is a sum of projectors defined as
\begin{equation}\label{eqn:proj}
P_0= \sum_{j=1}^n \ketbra{1}{1}_{j} = \frac{n}{2}\left(\eye - \frac{1}{n}\sum_{j=1}^n   Z_{j} \right).  
\end{equation} 
\end{definition}

To incorporate the gate sequence we consider \eqref{eqn:proj} as the initial Hamiltonian, with low energy state $\ket{0^n}$. We will act on \eqref{eqn:proj} with a sequence of gates $\prod_{l=1}^p U_l$ corresponding to the circuit being simulated as 
\begin{equation}\label{eqn:isoaffine}
h(k) = \left(\prod_{l=1}^{k\leq p} U_l \right)P_0 \left(\prod_{l=1}^{k\leq p} U_l\right)^\dagger \geq 0
\end{equation}
which is isospectral with \eqref{eqn:proj}.\footnote{I.e.~$P_0 \ket{x} = |x|_1\ket{x}$ for $x\in \{0,1\}^{\times n}$ and $|\cdot|_1$ the Hamming weight.} It follows that $h(k)$ is non-negative $\forall 1\leq k \leq p$ and that:  
\begin{equation}
    \ker\{h(p)\}=\text{span}\{\Pi_{l=1}^p U_l \ket{0^n} \}. 
\end{equation}

\begin{proposition}[Clifford Penalty Functions~\cite{Bia21}] \label{prop:pen} 
Let $\Pi_{l=1}^p U_l \ket{0^n}$ be a $p$-gate quantum circuit preparing state $\ket{\psi}$ on $n$-qubits and containing $\mathcal{O}(\text{poly}(\ln n))$ non-Clifford gates. Then there exists a non-negative Hamiltonian $H$ on n-qubits with $\text{poly}(p, n)$ cardinality, gap $\Delta$ and $\ker\{ H \}= \text{span}\{ \Pi_{l=1}^p U_l \ket{0^n}\}$. In particular, if $\ket{\phi}$ is such that 
\begin{equation}
0\leq \bra{\phi} H \ket{\phi}  < \Delta 
\end{equation}  it follows that 
\begin{equation}\label{eqn:stabs}
1 - \frac{\bra{\phi} H\ket{\phi} }{\Delta} \leq | \braket{\phi}{\psi}|^2 \leq 1 - \frac{\bra{\phi}H\ket{\phi} }{\max\{ H\}}.
\end{equation} 
\end{proposition}

The Hamiltonian described in Proposition \ref{prop:pen} is given exactly in Definition \ref{def:telescopes}.  To understand the utility of this, let us recall the following: 

\begin{definition}[Operator cardinality~\cite{Bia21}]
Let $H=\sum_k h_k\bigotimes_{j=1}^{n} \sigma_j^{\alpha_{j}(k)}$ for coefficients $h_k$ and Pauli strings $\bigotimes_{j=1}^{n} \sigma_j^{\alpha_{j}(k)}$.  Then $|H~|_{\text{card}}= \sum_k (h_k)^0$. 
\end{definition}

We then note that the telescopic construction gives rise to an $n$-term penalty function

\begin{lemma}[Clifford Gate Cardinality Invariance~\cite{Bia21}] \label{lemma:invariance}
For $C$ a Clifford gate and $h\in \text{span}_\mathbb{R}\left\{ \bigotimes_{l=1} \sigma^{\alpha_{l}}_l  \mid \alpha_{l} = 0,1,2,3 \right\}$, $|h|_{\text{card}} = |C h C^\dagger|_{\text{card}}$. 
\end{lemma}










\begin{example}[Parent Hamiltonian of the GHZ state]
The 3-qubit GHZ state is given as:
\begin{equation}\label{eqn:GHZ}
    \ket{\textrm{GHZ}}=G\ket{000}= \frac{1}{\sqrt{2}}\left(\ket{000} + \ket{111} \right),
\end{equation}
where $G = (H\otimes \eye \otimes \eye)\cdot (C(X)\otimes \eye) \cdot (\eye \otimes C(X))$. Graphically \eqref{eqn:GHZ} reads as:

\begin{equation*}
    \includegraphics[width=.3\textwidth]{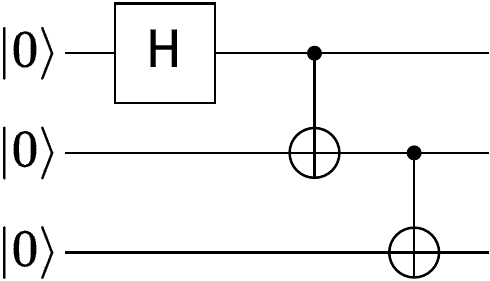}
\end{equation*}
Using the telescoping construction, Definition \ref{def:telescopes}, we write the Hamiltonian that has the GHZ state as its ground state as: $H_{\textrm{GHZ}}= G P_0 G^\dagger$. 
Graphically it reads as:

\begin{equation*}
    \includegraphics[width=.7\textwidth]{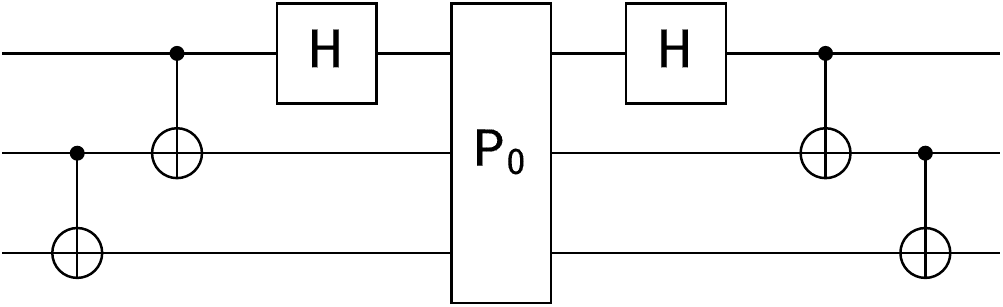}
\end{equation*}

We retrieve $H_{\textrm{GHZ}}$ by doing the following graphical calculations:

\begin{equation}\label{eqn:GHZcalc}
    \includegraphics[width=\textwidth]{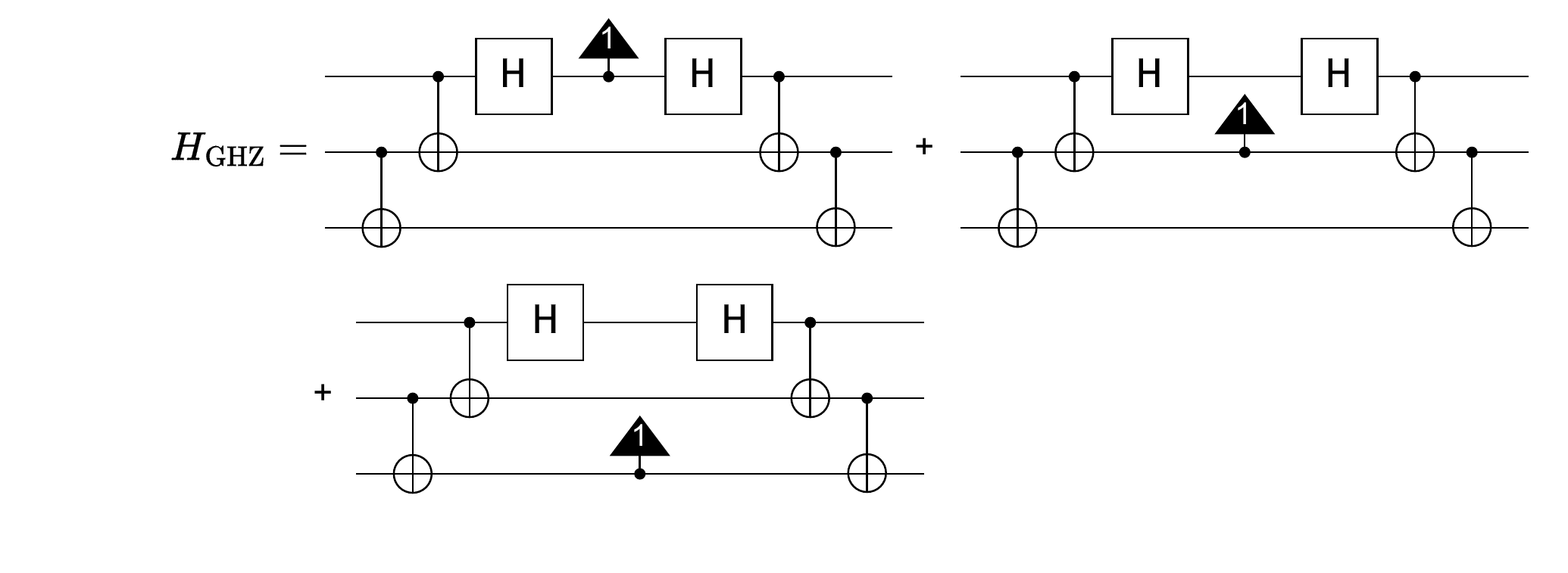}
\end{equation}
Using the Heisenberg rules the first term in \eqref{eqn:GHZcalc} is rewritten as:

\begin{equation}\label{eqn:GHZfirstterm}
    \includegraphics[width=\textwidth]{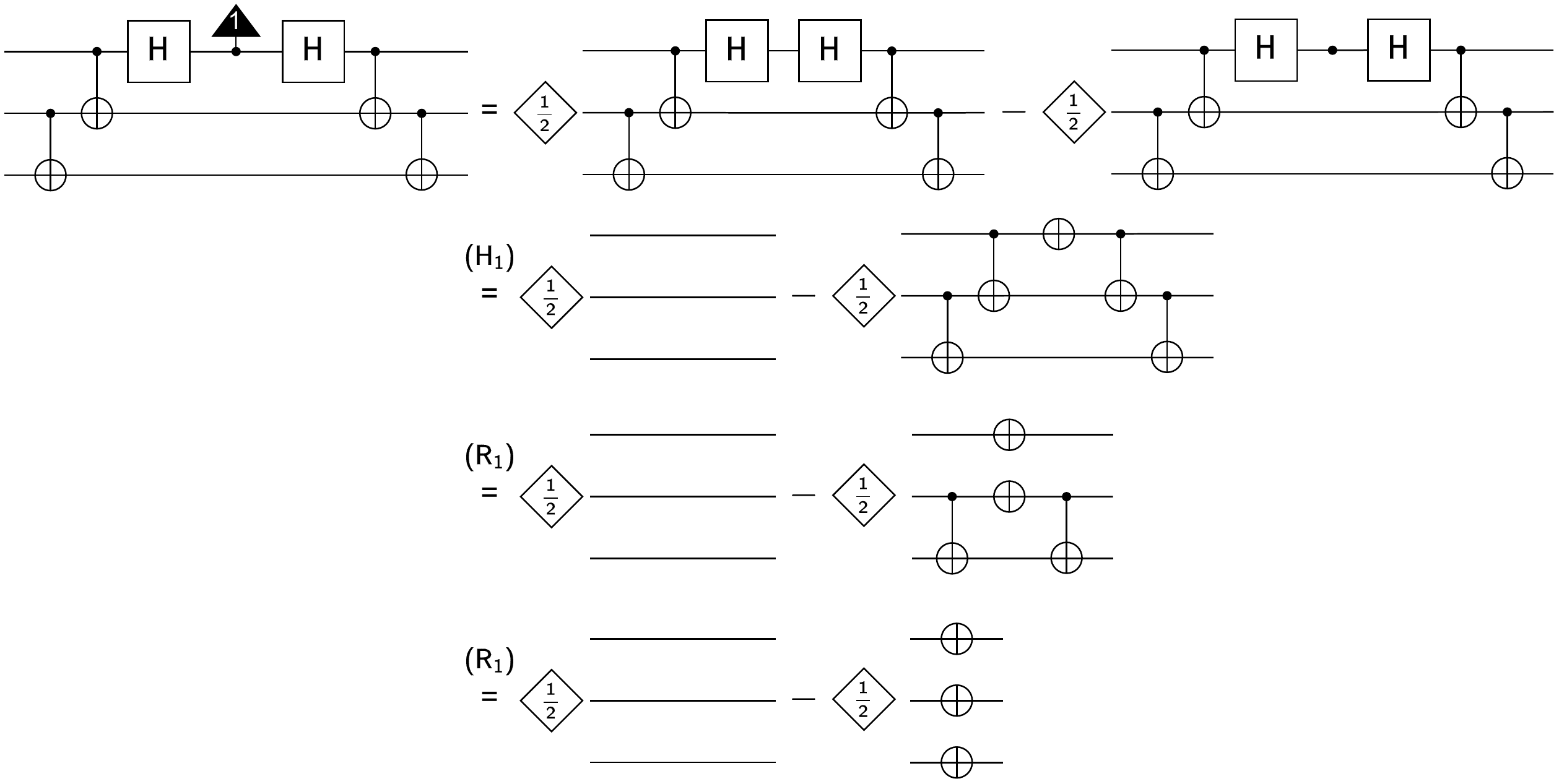}
\end{equation}
which is equal to, $\frac{1}{2}\eye - \frac{1}{2}(X+X+X)$. Finally, following the same calculation of \eqref{eqn:GHZfirstterm} we get,

\begin{equation}
    H_{\textrm{GHZ}}= \frac{3}{2} \eye- \frac{1}{2}  (X \otimes X \otimes X+ Z \otimes Z\otimes \eye+ \eye \otimes Z\otimes Z)
\end{equation}
This expression can be generalized for arbitrary $n$ qubits:
\begin{equation}\label{eqn:ghzn} 
    H_{\textrm{GHZ}}^{(n)} = \frac{1}{2} \left(1- \bigotimes_{j=1}^n X_j\right) + \frac{1}{2}\sum_{j=1}^n \left(1-  Z_j Z_{j+1}\right).
\end{equation}
where $j$ is the qubit index.
\end{example}

As a penalty function, \eqref{eqn:ghzn} is non-negative and has a one-dimensional kernel spanned by the $n$-qubit GHZ state.  Returning back to Proposition \ref{prop:pen}, the circuit which prepares the GHZ state defines the parent Hamiltonian \eqref{eqn:ghzn}.

\section{Conclusion}\label{conclusion}

Soundness underpins any graphical language, be it rudimentary template rewrite systems~\cite{Mas+05,Bra+21,RDH14} or the  ZX-calculus \cite{CD,redgreen}. In general, whereas every template is proven by matrix equality, not every template can be derived graphically.  Indeed, the ZX-calculus is generally known to be incomplete \cite{WZ14}. Furthermore, the ZX-calculus remains incomplete for the Clifford+T subset of quantum mechanics \cite{Jea+17}, it is however complete for certain restrictions and modifications.  Completeness has been shown for stabiliser quantum mechanics \cite{Backens_2014}, real-valued stabiliser quantum mechanics \cite{DP13} and when representing some general matrix equations \cite{JPV17}.  Conceptually one can understand completeness and soundness as a disjoint partitioning of ZX-diagrams.  Each disjoint set is related under ZX-equivalent rewrites.  

With these results established, little attention has been given to the confluence of restricted forms of the ZX-system.  Interestingly, the  stabilizer ZX-rewrite system is confluent and also terminal in the Heisenberg picture.  This a strong condition, that the ZX-system is Heisenberg-canonical.  The next steps include universal completion by considering the addition of T-gates \cite{JPV17} as well as further derivation of Hamiltonian penalty functions (such as the telescopes \cite{Bia21} for use in variational quantum computation) graphically.  One can conceptualize these results as partitioning Heisenberg ZX-diagrams into a disjoint set of directed graphs, each terminating at a Pauli term under Heisenberg ZX-rewrites.

\section{Acknowledgements}
AN acknowledges support from the research project No.~014/20, \textit{Leading Research Center: Quantum Computing}.  The authors thank Soumik Adhikary and Richik Sengputa for feedback.  

\bibliographystyle{unsrt}
 \bibliography{refs}


\appendix

\section{Heisenberg Rules}\label{appendix:Heisenberg-rules}
\S~\ref{sec:templates} shows how the generators of the Pauli group evolve under the conjugation of Clifford gates. In this section, we include the $Y$ gate and derive the remaining Heisenberg templates.The conjugation of the $Y$ gate under $H$, and $P$ is as follows:
\begin{equation}\label{eqn:Yconj}
  H\colon Y \rightarrow -Y, ~~~~  P\colon  Y \rightarrow -X.
\end{equation}
The l.h.s~of \eqref{eqn:Yconj} is given graphically as:
\begin{equation}
    \includegraphics[width=.5\textwidth]{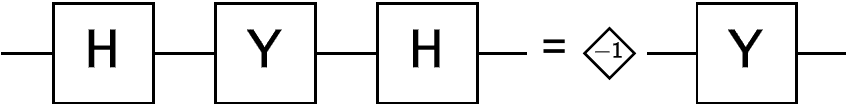}
\end{equation}
and the r.h.s is given as:
\begin{equation}
    \includegraphics[width=.5\textwidth]{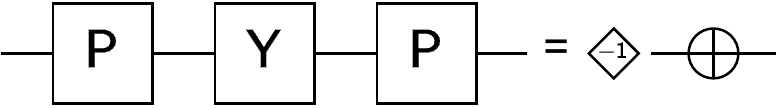}
\end{equation}

The remaining Heisenberg templates are derived graphically as:
\paragraph{(i). $C(X)\colon Y\otimes \eye \rightarrow Y\otimes X$}
\begin{equation}
    \includegraphics[width=.9\textwidth]{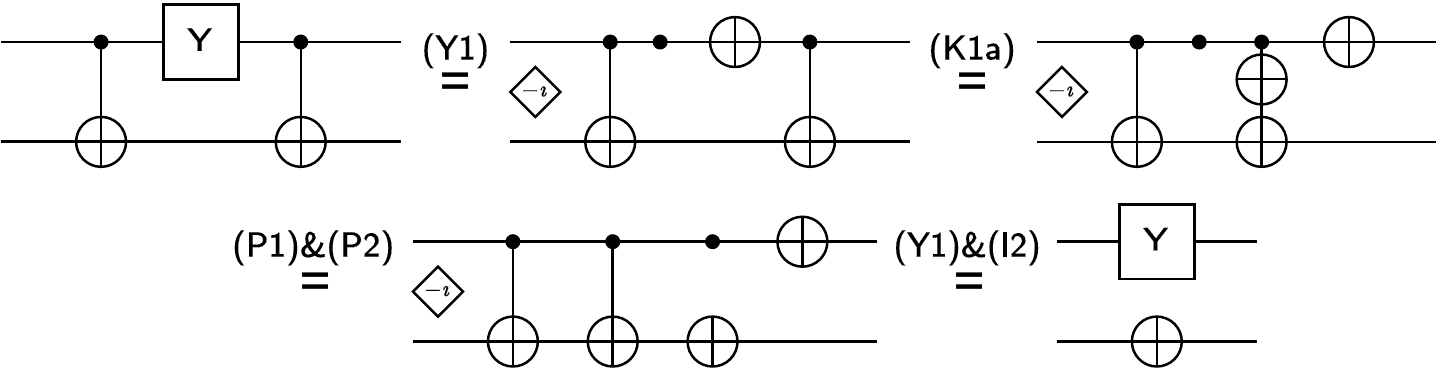}
\end{equation}

\paragraph{(ii). $C(X)\colon \eye\otimes Y \rightarrow Z\otimes Y$}
\begin{equation}
    \includegraphics[width=.9\textwidth]{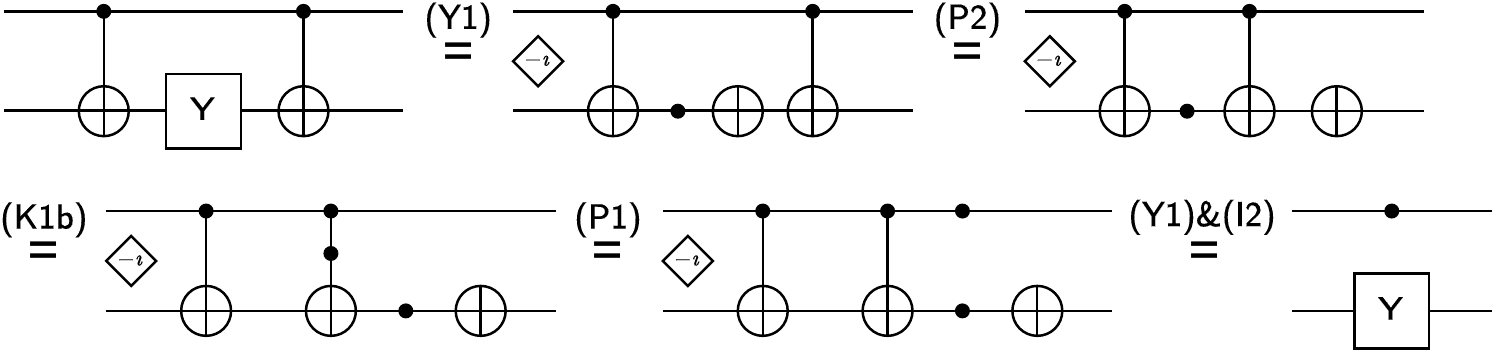}
\end{equation}

\paragraph{(iii). $C(X)\colon Y\otimes Y\rightarrow -X\otimes Z$}
\begin{equation}
    \includegraphics[width=.9\textwidth]{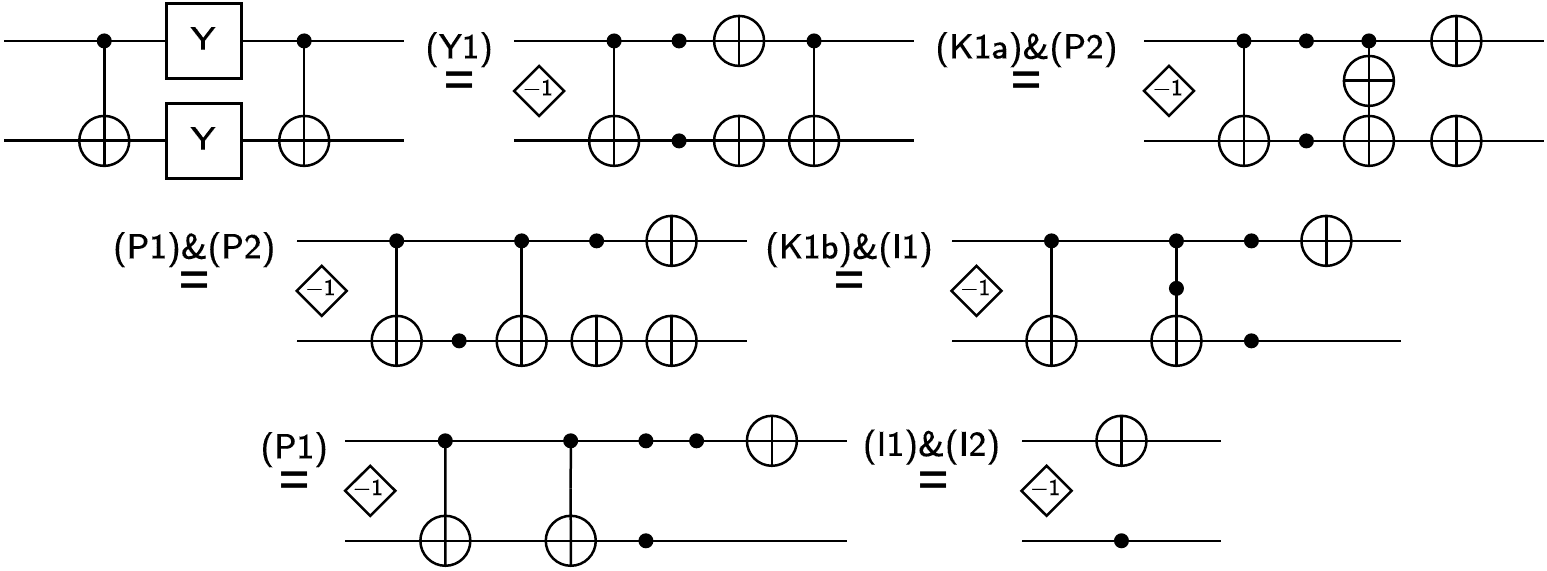}
\end{equation}

\paragraph{(iv). $C(X)\colon X\otimes Y  \rightarrow Y\otimes Z$}
\begin{equation}
    \includegraphics[width=.9\textwidth]{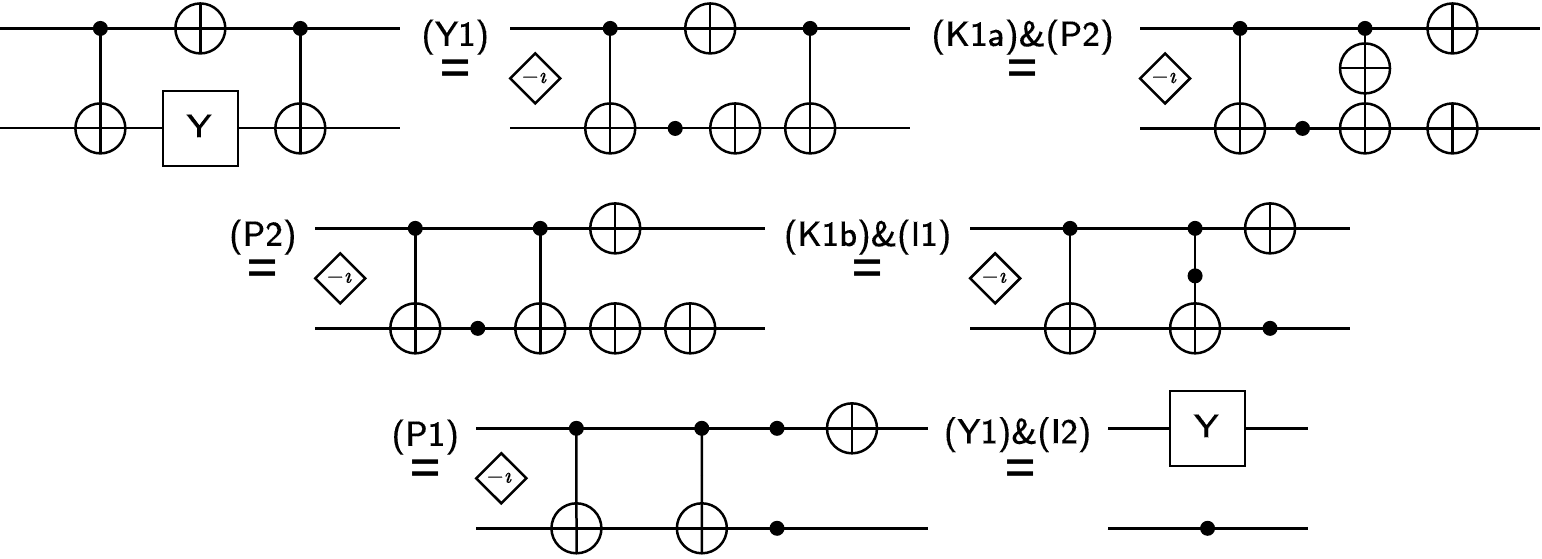}
\end{equation}

\paragraph{(v). $C(X)\colon Y\otimes X  \rightarrow Y\otimes \eye$}
\begin{equation}
    \includegraphics[width=.9\textwidth]{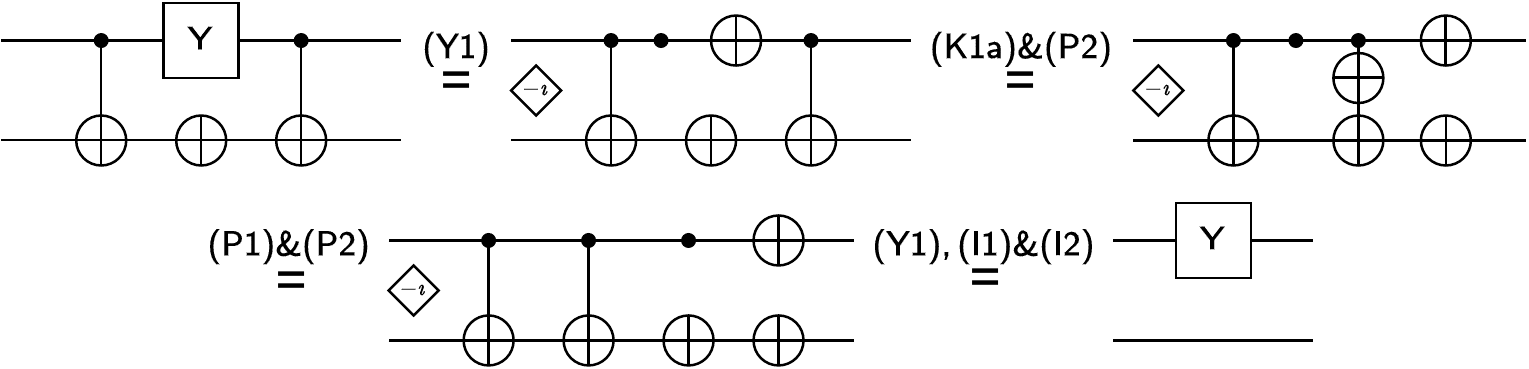}
\end{equation}

\paragraph{(vi). $C(X)\colon Y\otimes Z  \rightarrow X\otimes Y$}
\begin{equation}
    \includegraphics[width=.9\textwidth]{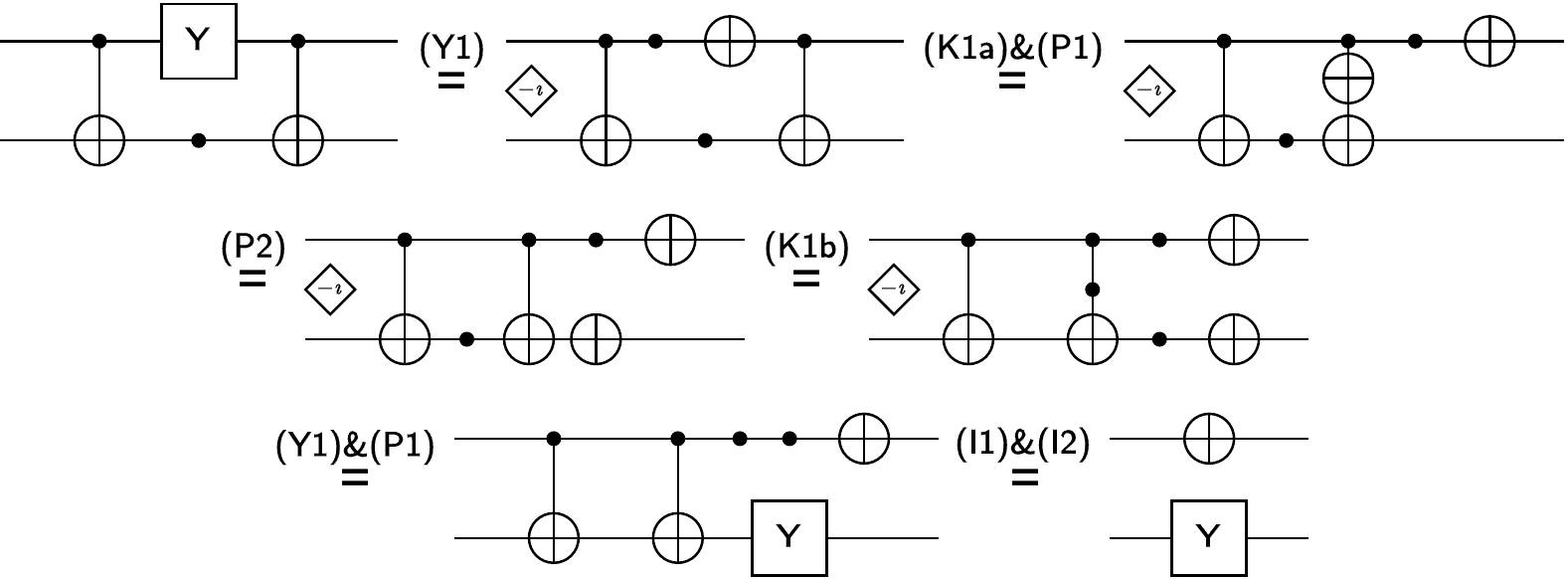}
\end{equation}

\paragraph{(vii). $C(X)\colon Z\otimes Y  \rightarrow \eye \otimes Y$}
\begin{equation}
    \includegraphics[width=.9\textwidth]{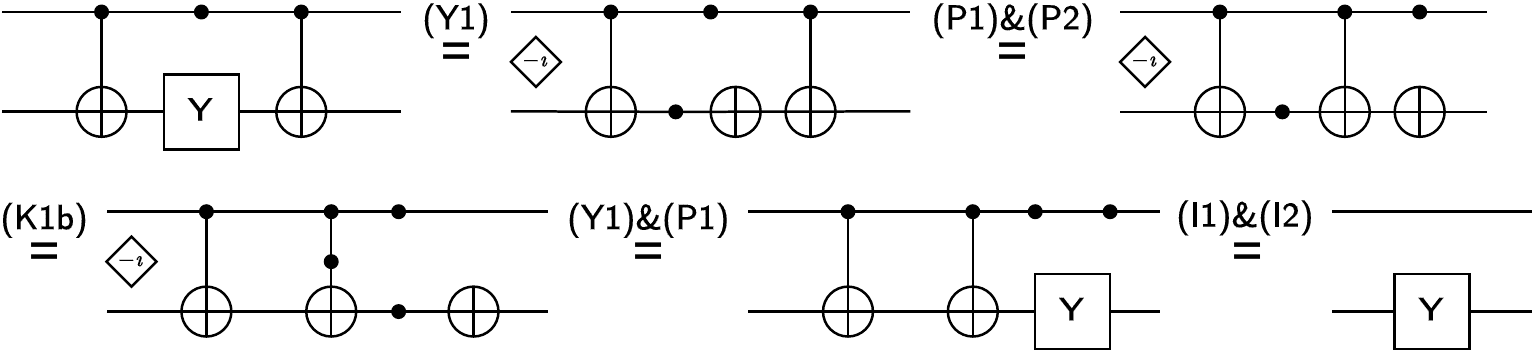}
\end{equation}

\end{document}